# A component-level co-rotational 3D continuum finite element framework for efficient flexible multibody analysis

Ziyun Kan[1]*, Mingdong Chen[2], Haijun Peng[2], Yizhu Guo[3], Xueguan Song[1]

1 State Key Laboratory of High-performance Precision Manufacturing, School of Mechanical Engineering, Dalian University of Technology, Dalian 116024, People's Republic of China.

2 State Key Laboratory of Structural Analysis for Industrial Equipment, Department of Engineering Mechanics, Dalian University of Technology, Dalian 116024, People 's Republic of China.

3 Beijing Institute of Spacecraft System Engineering, Beijing 100094, People 's Republic of China.

**Abstract:**

One of the most worrisome issue in flexible multibody dynamic analysis is its computational efficiency, since very often large-scale dimensions are involved in the finite element discretization. The most time-consuming process in traditional implicit integration analysis is updating the Jacobian matrices and solving the large-scale linearized equations. This paper proposes a systematic and novel framework, namely component level co-rotational (CR) approach, for upgrading existing 3D continuum finite elements to flexible multibody analysis. A significant innovation of the proposed framework is its high efficient for implicit analysis even with a relatively large number of finite elements. The framework doesn't rely on using any model reduction techniques, and the high efficiency is achieved through sophisticated operations in both modeling and numerical implementation phrases. In modeling phrase, as in conventional 3D nonlinear finite analysis, the nodal absolute coordinates are used as the system generalized coordinates, therefore simple formulations of the inertia force terms can be obtained. For the elastic force terms, inspired by existing floating frame of reference formulation (FFRF) and conventional element-level CR formulation, a component-level CR modeling strategy is developed. In the numerical implementation phrase, by in combination with Schur complement theory and fully exploring the nature of the component-level CR modeling method, an extremely efficient procedure is developed, which enables us to transform the linear equations raised from each Newton-Raphson iteration step into linear systems where the coefficient matrix remains constant. The coefficient matrix thus can be pre-calculated and decomposed only once, and at all the subsequent time steps only back substitutions are needed, which avoids frequently updating the Jacobian matrix and avoids directly solving the large-scale linearized equation in each iteration. Multiple examples are presented to demonstrate the performance of the proposed framework. It is found that the results have a good agreement with those obtained by the conventional CR analysis as well as by the commercial software. The outcomes also show that with the proposed framework the computational efficiency can achieve 10~100 faster than the conventional CR analysis for the numerical examples tested in this work.

---

*Corresponding author. E-mail addresses: kanziyun@dlut.edu.cn ;

## 1. Introduction

Flexible multibody system plays an important role in concurrent computer aided technical mechanics. Major areas of application include vehicle dynamics, manipulators and robots, bio-dynamical systems, heavy machinery. Motivated by these applications, flexible multibody analysis has been the focus of intensive research for the last forty years (Bauchau, 2010; Eberhard and Schiehlen, 2006; Jalon and Bayo, 2011; Kan et al., 2021b; Kerst et al., 2018; Liang et al., 2018; Rong et al., 2018; Shabana, 1997; Wasfy and Noor, 2003b). Enhancing the computational efficiency, analyzing accuracy and modeling convenience are always the pursuit for developing new techniques or formulations for flexible multibody analysis.

The existing formulations for flexible multibody dynamic analysis can be roughly categorized into the following two types. The first one is reference coordinate-based method. The well-established floating frame of reference formulation (FFRF) (Nada et al., 2009; Shabana, 2005) belongs to this kind of method. In this method, a local frame attached to a flexible body is explicitly defined to reflect the body large overall motion. Deformations are measured with respect to this local frame through additional elastic coordinates which can be introduced using finite element discretization. This kind of method is historically an extension of the Cartesian formulation used in rigid body system analysis (Ellenbroek and Schilder, 2018) and it is widely used in flexible multibody dynamics literature. The advantage of the approach is that the elastic force can be easily formulated with a simple linear relationship with the elastic coordinates. While the disadvantage is that complicated embroil motions is involved in the kinematic relationship, leading to complex inertia force terms, i.e. highly nonlinear mass matrix, Coriolis and centrifugal forces.

The second type is the absolute coordinate-based formulation. The absolute nodal coordinate formulation (ANCF) (Escalona et al., 1998; Gerstmayr and Shabana, 2006; Shabana, 1998; Shabana et al., 1998) is a typical one of this kind. In this formulation, the nodal absolute coordinates and global slope vectors are adopted as element generalized coordinates. In addition to ANCF, the total/updated Lagrangian formulations (Bathe, 1996; Zienkiewicz et al., 2013) used in ordinary nonlinear 3D finite element analysis can also be viewed as this kind of method, since only nodal global displacements/positions are involved in the analysis and they are essentially the generalized coordinates. In contrast to FFRF, the displacement/position, velocity and acceleration of a particular point in this formulation can be written as a linear relationship with the generalized coordinates by using the shape functions, thus a simple formulation of kinematic relationship can be obtained. This lead to a constant mass matrix and no Coriolis and centrifugal term appears in the analysis. The price to pay is that the complexity transforms into the elastic force calculation. Due to large rotations it is necessary to use nonlinear strain-displacement relationships.

For flexible multibody system, it is usually the case that the bodies are subjected to large rotations but small strains. To effectively address the elastic force calculation in this situation, the co-rotational (CR) approach (Cho et al., 2017; Crisfield, 1990; Crisfield and Moita, 1996; Felippa and Haugen, 2005; Lesiv and Izzuddin, 2023; Nour-Omid and Rankin, 1991; Rankin and Brogan, 1986; Rankin and Nour-Omid, 1988) can be adopted as a simple way alternative to conventional Lagrangian formulations. The main concept of CR approach is to decompose the motion of the element into rigid-body motion and the deformational parts, through the use of a local reference frame which continuously rotates with the element. The deformational response is captured at the level of the element local frame, whereas the geometric nonlinearity induced by the large rigid-body motion is incorporated in the transformation matrices relating local and global quantities. The idea of CR approach is somewhat similar to FFRF if we consider the local frame as the "floating frame" in FFRF, but two significant differences are worthy noting. The first one is that in CR approach the information of the local reference frame is calculated based on the nodal absolute coordinates using some direct or indirect rules, while in FFRF the information of floating frame is directly reflected through the generalized coordinates. The second difference is that in FFRF the elements located at the same flexible body shares a common reference frame, while in conventional CR approach, each element possesses an individual reference frame.

With regards to numerical calculation, the dynamic equation of flexible multibody systems is typically a set of highly nonlinear stiff differential algebraic equations (DAEs), and implicit integration algorithms are usually needed. The algorithms discretize the DAEs into a set of nonlinear algebraic equations, of which the solution can be obtained by classical Newton-Raphson method. The key process in the analyses is updating the Jacobian metrics and solving the large-scale linearized equations, which requires a large amount of computational resources. To enhance the computational efficiency, model reduction techniques are commonly adopted in existing formulations. For FFRF, this is popularly done using component mode synthesis method (Acri et al., 2016; Cammarata and Pappalardo, 2020; Held et al., 2015; Lozovskiy, 2014; Wasfy and Noor, 2003a; Ziegler et al., 2016). Although this technique works well and is extensively used, it might suffer from complexities, i.e., highly nonlinear inertia terms, the requirements of pre modal analysis and careful attention of the number of truncated mode. For ANCF, it is typically more difficult to conduct model reduction given the highly nonlinear relationship between the elastic force and the generalized coordinates. Some hyper reduction-based techniques (Hou et al., 2020) have been recently employed in the analysis, which is data-driven (Peng et al., 2022) and the effects heavily depend on the data integrity before the reduction.

Inspired by existing formulations in flexible multibody analysis and geometric nonlinear finite element analysis, the objective of this paper is to propose a systematic and novel framework, namely component-level CR approach, for upgrading existing 3D finite elements to allow flexible multibody analysis. A significant innovation of the proposed framework is its high efficient for implicit analysis even with a relatively large number of finite elements and without using any model reduction techniques. The proposed framework differs from the existing well-established FFRF and only the absolute nodal coordinates are adopted. As a result, the framework inherits the advantages of using the absolute coordinate-based formulations, i.e., simple formulations of inertia force terms, constant mass matrix (the same matrix that appears in linear structural dynamics) and avoidance of complicated Coriolis and centrifugal force terms. For elastic force calculations, inspired by the idea of conventional CR approach and FFRF, a component-level CR modeling approach is developed. By noting that the rotational invariance property of the mass matrix and fully exploring the nature of the component-level CR approach, an extremely efficient procedure for numerical solving the governing equation is developed, in the context of classical implicit Newmark algorithm. The rest of this paper is organized as follows. Section 2 presents a briefly review and discussion on the concept of conventional CR formulation used in geometric nonlinear finite element analysis. Section 3 gives the modeling method of the proposed framework for flexible multibody analysis. Effective numerical implementations that are tailored for the proposed modeling approach is given in Section 4. To promote applications, some discussions are presented in Section 5. Multiple examples are given in Section 6 to demonstrate the numerical performances of the proposed framework. Finally, some conclusions are summarized in Section 7.

## 2. Conventional CR formulation

The CR approach is viewed as a simple way to conduct nonlinear finite element analysis for large displacement but small strain problems. To present a comprehensive insight into the framework to be proposed, we first briefly review and discuss the concept of conventional CR formulation. Throughout this work, we limit our discussion to 3D continuum elements with small strain.

We first focus on elastic internal force terms. Consider a general 3D continuum element with $N$ nodes shown in Fig. 1. Each node consists of three translational degrees of freedom. $OXYZ$ denotes the global coordinate frame. The element stiffness matrix associated with initial frame is denoted as $\bar{\boldsymbol{K}} \in \Re^{3N\times 3N}$, which is a symmetric positive semidefinite matrix. The element stiffness matrix has the property that for arbitrary element rigid-body translation $\bar{\boldsymbol{u}}_{\mathrm{tran}} \in \Re^{3N\times 3N}$, no internal force will be produced (i.e., $\bar{\boldsymbol{K}}\bar{\boldsymbol{u}}_{\mathrm{tran}} \equiv \boldsymbol{0}$). If the element is subjected to a small nodal displacement $\boldsymbol{u}$ such that the element rotation is small, the strain energy can be given as $\Phi(\boldsymbol{x}) = \boldsymbol{u}^{\mathrm{T}}\bar{\boldsymbol{K}}\boldsymbol{u}/2$. Trivially, by taking the variation of the strain energy to the displacement vector the nodal internal force vector can be obtained as $\boldsymbol{f}_{\mathrm{INT}} = \left[\mathrm{d}\Phi/\mathrm{d}\boldsymbol{u}\right]^{\mathrm{T}} = \bar{\boldsymbol{K}}\boldsymbol{u}$

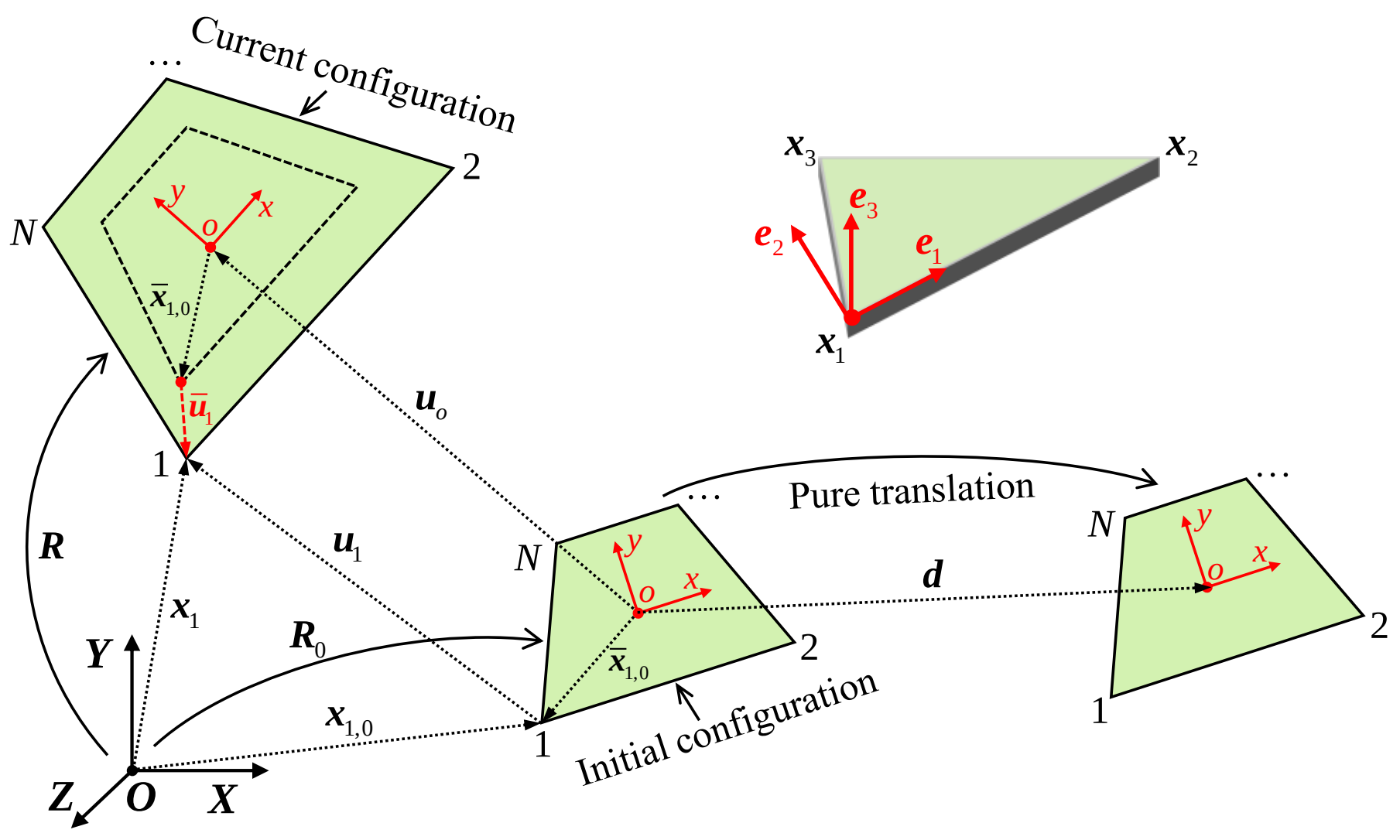

Fig. 1 CR formulation of a continuum element: kinematics and coordinate systems.

We now consider the case that element is subjected to a large nodal displacement. The main idea of CR approach is to extract the local deformation part from the large displacement, such that the strain energy function can still be approximated at the small deformation region. This is done through defining a local CR coordinate system and measuring all deformations in this local CR frame. With the help of element local CR frame, the element kinematics can be split into two steps. The first step is rigid translation and rotation of the local CR frame, and the second one is local deformation with respect to this local CR frame. The element local deformational displacement, $\bar{\boldsymbol{u}} \in \Re^{3N}$, is typically given as

$$\bar{\boldsymbol{u}} = \mathrm{diag}(\boldsymbol{R})^{\mathrm{T}} \left(\boldsymbol{u} + \boldsymbol{x}_0 - (\boldsymbol{u}_{\mathrm{o}} + \boldsymbol{x}_{\mathrm{o},0})\right) - \bar{\boldsymbol{x}}_0 \tag{1}$$

where $\boldsymbol{u} \in \Re^{3N}$ and $\boldsymbol{x}_0 \in \Re^{3N}$ denote the global displacement and initial position of the nodes, respectively; $\bar{\boldsymbol{x}}_0 \in \Re^{3N}$ denotes the initial nodal position vector measured in the local CR frame, and it is a constant throughout the analysis; $\boldsymbol{u}_{\mathrm{o}} \in \Re^{3N}$ and $\boldsymbol{x}_{\mathrm{o},0} \in \Re^{3N}$ are the column vector arrayed by the global displacement and initial position of the origin of the local CR frame; $\mathrm{diag}(\boldsymbol{R})$ denotes a $3N\times3N$ block diagonal matrix composed of the orthogonal matrix $\boldsymbol{R} \in \Re^{3\times3}$:

$$\mathrm{diag}(\boldsymbol{R}) = \begin{bmatrix} \boldsymbol{R} & & & \\ & \boldsymbol{R} & & \\ & & \ddots & \\ & & & \boldsymbol{R} \end{bmatrix} \tag{2}$$

Expression (1) represents the general formulation of local deformation in many existing literature (Felippa and Haugen, 2005; Nour-Omid and Rankin, 1991; Rankin and Brogan, 1986; Rankin and Nour-Omid, 1988). However, this expression can be simplified by dropping terms $\boldsymbol{u}_0$ and $\boldsymbol{x}_{0,0}$, since these terms only influence element rigid-body translation and have no contribution to the internal force calculation. Therefore, we have

$$\bar{\boldsymbol{u}} = \mathrm{diag}(\boldsymbol{R})^{\mathrm{T}} \boldsymbol{x} - \bar{\boldsymbol{x}}_0 \tag{3}$$

where $\boldsymbol{x} = \boldsymbol{u} + \boldsymbol{x}_0 \in \Re^{3N}$ means the current position vector of the element nodes. The orthogonal matrix $\boldsymbol{R}$ represents the rigid rotation of the element local CR frame, which is evaluated using some geometry-based or polar decomposition-based rules according to the current configuration $\boldsymbol{x}$. One commonly used geometry-based rule is the three-node side alignment approach. Taking the nodes 1-2-3 as an example, the current element rotational matrix $\boldsymbol{R}(\boldsymbol{x}) = [\boldsymbol{e}_1, \boldsymbol{e}_2, \boldsymbol{e}_3] \in \Re^{3\times 3}$ can be determined with (as indicated in the upper right of Fig. 1)

$$\boldsymbol{e}_1 = \frac{\boldsymbol{x}_2 - \boldsymbol{x}_1}{\|\boldsymbol{x}_2 - \boldsymbol{x}_1\|}; \boldsymbol{e}_3 = \frac{(\boldsymbol{x}_2 - \boldsymbol{x}_1)\times(\boldsymbol{x}_3 - \boldsymbol{x}_1)}{\|(\boldsymbol{x}_2 - \boldsymbol{x}_1)\times(\boldsymbol{x}_3 - \boldsymbol{x}_1)\|}; \boldsymbol{e}_2 = \boldsymbol{e}_3 \times \boldsymbol{e}_1 \tag{4}$$

This rule will be used throughout this work given its simplicity and efficiency.

Based on the core concept of CR, the strain energy function is obtained as

$$\Phi(\boldsymbol{x}) = \bar{\boldsymbol{u}}^{\mathrm{T}} \bar{\boldsymbol{K}} \bar{\boldsymbol{u}} / 2 \tag{5}$$

The element (global) internal force can thus be derived as

$$\boldsymbol{f}_{\mathrm{INT}}(\boldsymbol{x}) = \left[\frac{\mathrm{d}\Phi}{\mathrm{d}\boldsymbol{x}}\right]^{\mathrm{T}} = \left[\frac{\mathrm{d}\Phi}{\mathrm{d}\bar{\boldsymbol{u}}}\frac{\mathrm{d}\bar{\boldsymbol{u}}}{\mathrm{d}\boldsymbol{x}}\right]^{\mathrm{T}} = \left[\frac{\mathrm{d}\bar{\boldsymbol{u}}}{\mathrm{d}\boldsymbol{x}}\right]^{\mathrm{T}} \bar{\boldsymbol{K}}\bar{\boldsymbol{u}} \tag{6}$$

where the term $\bar{\boldsymbol{K}}\bar{\boldsymbol{u}}$ means the element local internal force, and $\mathrm{d}\bar{\boldsymbol{u}}/\mathrm{d}\boldsymbol{x}$ serves as a transformation matrix between the local force and the global one. Using Eq. (3) and after some derivations, this transformation matrix can be obtained (see in the Appendix)

$$\frac{\mathrm{d}\bar{\boldsymbol{u}}}{\mathrm{d}\boldsymbol{x}} = \mathrm{diag}(\boldsymbol{R})^{\mathrm{T}} + \left.\frac{\partial\left(\mathrm{diag}(\boldsymbol{R})^{\mathrm{T}}\boldsymbol{V}\right)}{\partial \boldsymbol{x}}\right|_{\boldsymbol{V}=\boldsymbol{x}} = \left(\boldsymbol{I}_{3N} - \bar{\boldsymbol{S}}\bar{\boldsymbol{G}}\right)\mathrm{diag}(\boldsymbol{R})^{\mathrm{T}} \tag{7}$$

where $\boldsymbol{I}_{3N}$ is $3N\times 3N$ identity matrix and $\bar{\boldsymbol{S}} \in \Re^{3N\times 3}$ is the spin-lever or moment-arm matrix (Felippa and Haugen, 2005) defined as

$$\bar{\boldsymbol{S}} = \left[ \text{spin}\left(\bar{\boldsymbol{x}}_1\right) \quad \text{spin}\left(\bar{\boldsymbol{x}}_2\right) \quad \cdots \quad \text{spin}\left(\bar{\boldsymbol{x}}_N\right) \right]^{\mathrm{T}} \tag{8}$$

where spin operation is given in the Appendix. Matrix $\bar{\boldsymbol{G}} \in \Re^{3\times 3N}$ in Eq. (7) is the spin-fitter matrix (Felippa and Haugen, 2005), defined as

$$\bar{\boldsymbol{G}} = \left[ \frac{\partial \bar{\boldsymbol{\omega}}}{\partial \bar{\boldsymbol{u}}_1} \quad \frac{\partial \bar{\boldsymbol{\omega}}}{\partial \bar{\boldsymbol{u}}_2} \quad \cdots \quad \frac{\partial \bar{\boldsymbol{\omega}}}{\partial \bar{\boldsymbol{u}}_N} \right] \tag{9}$$

where $\bar{\boldsymbol{\omega}}$ denotes the instantaneous axial vector of the orthogonal transformation expressed in the local CR frame. Detailed expression of matrix $\bar{\boldsymbol{G}}$ depends on how to approximate the orthogonal matrix $\boldsymbol{R}$ based on the current configuration, and it is typically a highly nonlinear function of the current quantities. Combining Eqs. (6)~(7), the element global internal force can be expressed as

$$\boldsymbol{f}_{\mathrm{INT}}\left(\boldsymbol{x}\right) = \text{diag}\left(\boldsymbol{R}\right)\bar{\boldsymbol{K}}\bar{\boldsymbol{u}} - \text{diag}\left(\boldsymbol{R}\right)\bar{\boldsymbol{G}}^{\mathrm{T}}\bar{\boldsymbol{S}}^{\mathrm{T}}\bar{\boldsymbol{K}}\bar{\boldsymbol{u}} \tag{10}$$

**Remark**: The above formulation represents the rigorous expression of the internal force. However, some issues must be noted to obtain a simpler formulation. As discussed in the authors' previous work (Kan et al., 2021a), the first term of the internal force, $\text{diag}\left(\boldsymbol{R}\right)\bar{\boldsymbol{K}}\bar{\boldsymbol{u}}$, represents the ordinary orthogonal transformation between the local force to the global one, and it is the *dominant part* of the internal force; the second term, $\text{diag}\left(\boldsymbol{R}\right)\bar{\boldsymbol{G}}^{\mathrm{T}}\bar{\boldsymbol{S}}^{\mathrm{T}}\bar{\boldsymbol{K}}\bar{\boldsymbol{u}}$, is a *high order small correction term* to the dominant part. The correction is only made for those degree of freedoms (DOFs) that are relevant to the determination of the orthogonal matrix. The effect of such a correction is to cure the possible error induced by the approximation of the element rigid rotation motion. If the element deformation is sufficiently small such that the element rigid rotation can be calculated quite precisely, there is completely no use to add the second term in the force calculation. The formulation of the internal force vector therefore can be simplified as

$$\boldsymbol{f}_{\mathrm{INT}}\left(\boldsymbol{x}\right) = \text{diag}\left(\boldsymbol{R}\right)\bar{\boldsymbol{K}}\bar{\boldsymbol{u}} \tag{11}$$

Other terms should be focused on in dynamic analysis are the inertia terms, such as the element mass matrix. For mass matrix, there is no need to follow a "CR" manner because it can be rigorously obtained from the kinetic energy of the element. The element kinetic energy is expressed by volume integral:

$$E = \frac{1}{2}\int_V \rho\left(\boldsymbol{N}_s\dot{\boldsymbol{x}}\right)^{\mathrm{T}} \boldsymbol{N}_s\dot{\boldsymbol{x}}dV = \frac{1}{2}\dot{\boldsymbol{x}}^{\mathrm{T}}\left(\int_V \rho\boldsymbol{N}_s^{\mathrm{T}}\boldsymbol{N}_s dV\right)\dot{\boldsymbol{x}} \tag{12}$$

where $\rho$ and $\boldsymbol{N}_s$ are the density and shape function of the element, respectively. As a result, the element mass matrix, known as consistent mass matrix, is obtained as

$$\boldsymbol{M}_{\mathrm{e}}=\int_{V}\rho\boldsymbol{N}_{s}^{\mathrm{T}}\boldsymbol{N}_{s}dV \tag{13}$$

The element mass matrix is a constant matrix and is the same as that appears in linear structural dynamics. Complicated Coriolis and centrifugal force terms is thus avoided according to the Lagrange's equation of motion. Another kind of mass matrix commonly used in engineering application is the lumped mass matrix, which is obtained by lumping the consistent mass matrix into a diagonal matrix.

**Remark**: Whether consistent mass matrix or lumped mass matrix, the matrix $\boldsymbol{M}_{\mathrm{e}}$ is a symmetric positive definite matrix and it has an important rotational invariance property, that is : for any given block orthogonal matrix the relationship $\mathrm{diag}(\boldsymbol{R})^{\mathrm{T}}\boldsymbol{M}_{\mathrm{e}}\mathrm{diag}(\boldsymbol{R})\equiv\boldsymbol{M}_{\mathrm{e}}$ holds. This property provides a foundation for our following effective numerical implementations of the proposed framework.

## 3. The proposed framework: modeling

The previous section reviews the conventional element-level CR formulation. This formulation can be directly adopted for flexible multibody analysis. This is down by meshing a flexible body into a certain number of finite elements, and each element can be straightforwardly analyzed using the previous formulation. However, since it is typically the case that a flexible body is meshed with a large amount of elements, the computational efficiency of such a direct method is usually unbearable, in particular for an analysis with tens of thousands of time steps. For implicit analysis, the computationally most expensive process is forming the large-scale current-state-dependent Jacobian matrix and solving the related linear equation at each Newton-Raphson iteration. In numerical examples of this paper, this kind of "full CR" approach will be conducted, with a moderate finite element mesh, in order to verify the solutions. To enhance the computational efficiency of CR approach in flexible multibody analysis, a component-level CR framework is proposed, as detailed below.

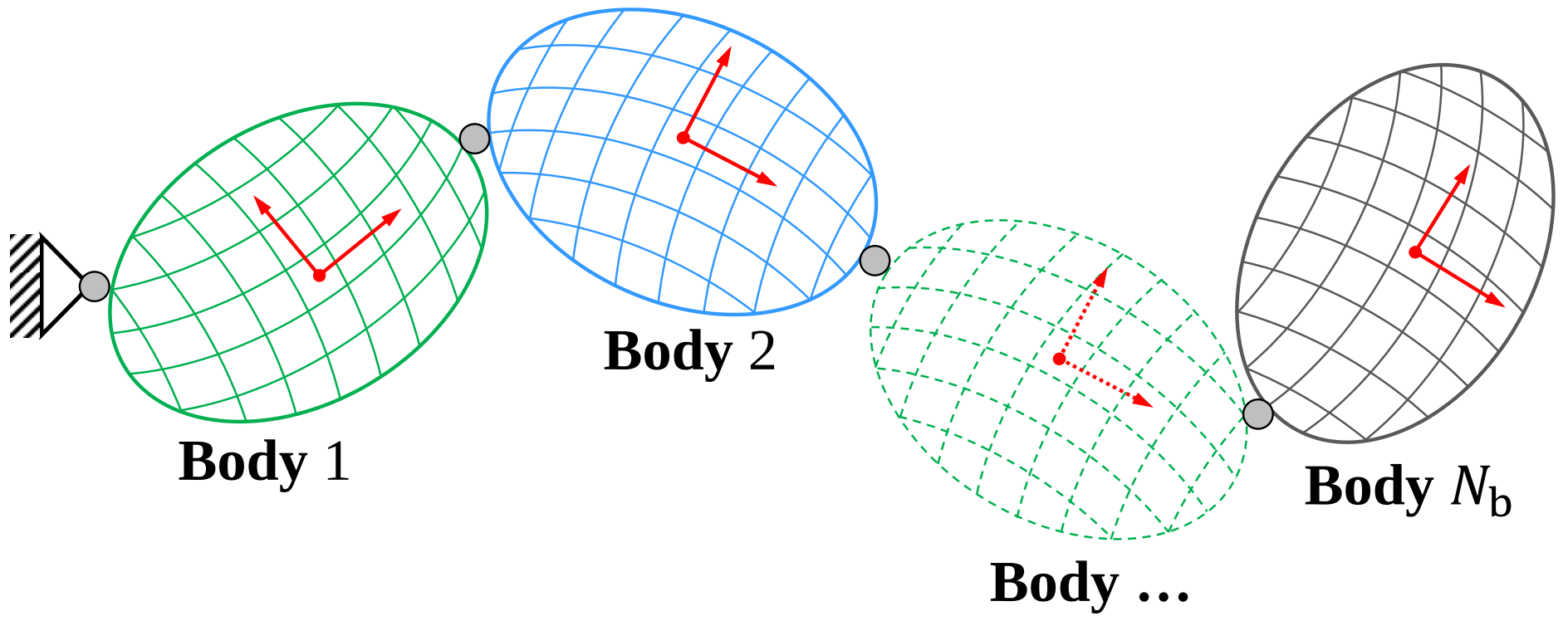

Fig. 2 A multibody body consists of $N_{\mathrm{b}}$ flexible bodies: each flexible is modeled using as a component, and the elements belongs to a flexible body share a single orthogonal matrix.

We consider a multibody body consists of $N_{\text{b}}$ flexible bodies shown in Fig. 2. All bodies are subjected to small deformations during the large overall motion. The basic idea of the proposed framework that we consider a flexible body as a component, and the elements belongs to a flexible body share a single orthogonal matrix, as is the case with FFRF. This orthogonal matrix characterizes the large overall motion of the flexible body. The generalized coordinate of each flexible body ( $\boldsymbol{x}_{\text{b}}^{(i)}$ ) is chosen as the nodal position vectors located on the body. Following a similar manner with the deduction of the internal force for an element, the internal force for the $i$th flexible body can be expressed as

$$\boldsymbol{f}_{\text{INT}}^{(i)}\left(\boldsymbol{x}_{\text{b}}^{(i)}\right)=\text{diag}\left(\boldsymbol{R}_{\text{b}}^{(i)}\right)\bar{\boldsymbol{K}}_{\text{b}}^{(i)}\bar{\boldsymbol{u}}_{\text{b}}^{(i)}=\text{diag}\left(\boldsymbol{R}_{\text{b}}^{(i)}\right)\bar{\boldsymbol{K}}_{\text{b}}^{(i)}\left(\text{diag}\left(\boldsymbol{R}_{\text{b}}^{(i)}\right)^{\text{T}}\boldsymbol{x}_{\text{b}}^{(i)}-\bar{\boldsymbol{x}}_{\text{b}0}^{(i)}\right) \tag{14}$$

where the terms with subscript "b" denotes that they are relevant to the entire flexible body. All terms in (14) are typically large-scale matrix/vectors, and the dimensions are dependent on the total DOF of the body. The mass matrix of the flexible body, $\bar{\boldsymbol{M}}_{\text{b}}^{(i)}$, can be obtained by assembling the mass matrix of each element belongs to the body, and the rotational invariance property $\text{diag}\left(\boldsymbol{R}_{\text{b}}^{(i)}\right)^{\text{T}}\bar{\boldsymbol{M}}_{\text{b}}^{(i)}\boldsymbol{R}_{\text{b}}^{(i)}\text{diag}\left(\boldsymbol{R}_{\text{b}}^{(i)}\right)\equiv\bar{\boldsymbol{M}}_{\text{b}}^{(i)}$ is retained. The orthogonal rotational matrix $\boldsymbol{R}_{\text{b}}^{(i)}$ is calculated based on the current nodal positions using some rules. In this work, the three-node side alignment approach illustrated in Eq. (4) is adopted given its simplicity. Theoretically, arbitrary non-collinear three nodes are practicable. To better reflect the mean movement of a body, we suggest the three nodes should be chosen such that they are far from each other.

In a multibody system, different bodies are usually connected through joints. Another important issue for the proposed framework is that we should ensure different flexible bodies are completely isolated in the finite element mesh. Taking the spherical joint as an example, in conventional finite element-based flexible multibody analysis, this kind of joint can be directly modeled by using a single node shared by two bodies. The related operation in calculation process is that some elements belong to the two bodies will both contribute to the DOFs of the shared node. The advantage of this strategy is that there is no need to introduce additional constraints to reflect the joint. However, in the proposed framework this strategy is not permissible, and the joint has to been modeled through additional constraint equations. By using isolated finite element mesh, we aim at ensuring that all terms related to different flexibles are non-overlapping, thus they can be addressed separately. This provides an important basis for developing effective solving algorithm.

The system generalized coordinate $\boldsymbol{q}$ is the sum set of the generalized coordinates of all the flexible bodies:

$$\boldsymbol{q}=\left[\boldsymbol{x}_{\text{b}}^{(1)\text{T}},\quad \boldsymbol{x}_{\text{b}}^{(2)\text{T}},\quad \cdots,\quad \boldsymbol{x}_{\text{b}}^{(N_{\text{b}})\text{T}}\right]^{\text{T}} \tag{15}$$

By employing the Lagrangian equation of motion and taking into account all the constraint equations, the system governing equation can be derived as the following index-3 differential algebraic equations (DAEs):

$$\begin{cases} \boldsymbol{M}\ddot{\boldsymbol{q}} + \boldsymbol{\Phi}_{\boldsymbol{q}}^{\mathrm{T}}\boldsymbol{\lambda} + \boldsymbol{f}_{\mathrm{INT}}\left(\boldsymbol{q}\right) = \boldsymbol{f}_{\mathrm{EXT}} \\ \boldsymbol{\Phi}\left(\boldsymbol{q}, t\right) = \boldsymbol{0} \end{cases} \tag{16}$$

where $\boldsymbol{M}$ is the system mass matrix, $\boldsymbol{\Phi}$ is the constraint equation, $\boldsymbol{\Phi}_{\boldsymbol{q}}$ is the constraint Jacobian matrix, $\boldsymbol{\lambda}$ is the Lagrange multiplier vector physically reflecting the constraint forces, $\boldsymbol{f}_{\mathrm{INT}}$ and $\boldsymbol{f}_{\mathrm{EXT}}$ are the internal force and the external force, respectively. The external force is considered independent of the system generalized coordinate (i.e., $\partial \boldsymbol{f}_{\mathrm{EXT}} / \partial \boldsymbol{q} = [\boldsymbol{0}]$). Coordinate-dependent external force will be discussed in Section 5.

The system mass matrix possesses a block diagonal structure,

$$\boldsymbol{M} = \begin{bmatrix} \bar{\boldsymbol{M}}_{\mathrm{b}}^{(1)} & & & \\ & \bar{\boldsymbol{M}}_{\mathrm{b}}^{(2)} & & \\ & & \ddots & \\ & & & \bar{\boldsymbol{M}}_{\mathrm{b}}^{(N_{\mathrm{b}})} \end{bmatrix} \tag{17}$$

and the internal force $\boldsymbol{f}_{\mathrm{INT}}$ possesses

$$\boldsymbol{f}_{\mathrm{INT}} = \left[ \boldsymbol{f}_{\mathrm{INT}}^{(1)\mathrm{T}}, \quad \boldsymbol{f}_{\mathrm{INT}}^{(2)\mathrm{T}}, \quad \cdots, \quad \boldsymbol{f}_{\mathrm{INT}}^{(N_{\mathrm{b}})\mathrm{T}} \right]^{\mathrm{T}} \tag{18}$$

Eq. (16) is a set of highly nonlinear, and typically stiff, differential algebraic equations. We adopt the classical implicit Newmark method to solve the problem. The method discretizes the motion equation into a set of nonlinear algebraic equations at each integral/time step. Provided that the solutions of the previous $n$th time step have been obtained, discretization of Eq. (16) at the $n$+1th time step and introduction of the Newmark assumptions yield the following nonlinear equations:

$$\begin{cases} \boldsymbol{M}\ddot{\boldsymbol{q}}_{n+1} + \boldsymbol{\Phi}_{\boldsymbol{q}_{n+1}}^{\mathrm{T}}\boldsymbol{\lambda}_{n+1} + \boldsymbol{f}_{\mathrm{INT}}\left(\boldsymbol{q}_{n+1}\right) = \boldsymbol{f}_{\mathrm{EXT}} \\ \boldsymbol{\Phi}\left(\boldsymbol{q}_{n+1}, t_{n+1}\right) / \alpha h^2 = \boldsymbol{0} \\ \dot{\boldsymbol{q}}_{n+1} = \dot{\boldsymbol{q}}_n + \left[ (1-\delta)\ddot{\boldsymbol{q}}_n + \delta \ddot{\boldsymbol{q}}_{n+1} \right] h \\ \boldsymbol{q}_{n+1} = \boldsymbol{q}_n + \dot{\boldsymbol{q}}_n h + \left[ (1-2\alpha)\ddot{\boldsymbol{q}}_n + 2\alpha \ddot{\boldsymbol{q}}_{n+1} \right] h^2/2 \end{cases} \tag{19}$$

where $h$ is the integration step size, and $\alpha$ and $\delta$ are the Newmark parameters. The constraint equations being multiplied by a scaling factor $1/\alpha h^2$ is to ensure the symmetry of the algorithm stability (Bottasso et al., 2007). Substituting the last two equations of (19) into the first two equations, the nonlinear equation is reduced to a system of $\boldsymbol{X} = \left[ \ddot{\boldsymbol{q}}_{n+1}^{\mathrm{T}}, \boldsymbol{\lambda}_{n+1}^{\mathrm{T}} \right]^{\mathrm{T}}$. The Newton-Raphson method can be applied to obtain the solution. At the $i$th iteration the linearized equation yields

$$\boldsymbol{J}\Delta\boldsymbol{X} = \boldsymbol{G}_{\mathrm{RES}} \tag{20}$$

where $\boldsymbol{J}$ is the Jacobian matrix given as

$$\boldsymbol{J} = \begin{bmatrix} \boldsymbol{M} + \alpha h^2 \dfrac{\partial\left(\boldsymbol{\Phi}_{\boldsymbol{q}_{n+1}}^{\mathrm{T}} \boldsymbol{\lambda}_{n+1}\right)}{\partial \boldsymbol{q}_{n+1}} + \alpha h^2 \dfrac{\partial \boldsymbol{f}_{\mathrm{INT}}}{\partial \boldsymbol{q}_{n+1}} & \boldsymbol{\Phi}_{\boldsymbol{q}_{n+1}}^{\mathrm{T}} \\ \boldsymbol{\Phi}_{\boldsymbol{q}_{n+1}} & \boldsymbol{0} \end{bmatrix} \tag{21}$$

and $\boldsymbol{G}_{\mathrm{RES}} = \left[\boldsymbol{F}^{\mathrm{T}}, \boldsymbol{\Psi}^{\mathrm{T}}\right]^{\mathrm{T}}$ denotes the residual error of the nonlinear equations; $\Delta\boldsymbol{X} = \left[\Delta\ddot{\boldsymbol{q}}_{n+1}^{\mathrm{T}}, \Delta\boldsymbol{\lambda}_{n+1}^{\mathrm{T}}\right]^{\mathrm{T}}$ denotes the iterative increment. Once system (20) is solved , The solution updates as $\boldsymbol{X} = \boldsymbol{X} - \Delta\boldsymbol{X}$ . This process repeats until the residual error is in some sense blow the given tolerance.

The above operation represents the ordinary implementation of the algorithm, and it works well for small-scale problems as both the calculation of Jacobian matrix (21) and the solving of linear system (20) can be done effectively. However, since it is typically the case that flexible multibody systems are meshed with a large amount of finite elements, the computational efficiency of such an ordinary implementation is usually unbearable. Even though some iterative-based strategies have been used in multibody analysis (Serban et al., 2015), the efficiency is still hard to meet the requirement for fast analyses. In next section we will develop an effective numerical implementation by fully exploring of the nature of the proposed component-level CR modeling method.

## 4. The proposed framework: effective numerical implementation

### 4.1 Simplification of the Jacobian matrix

Expression (21) represents the rigorous formulation of the Jacobian matrix. To seek an effective numerical implementation, the first step needed to do is simplifying the Jacobian matrix, which is done by dropping some unimportant terms. This step would not make influence on the solution of the nonlinear equation as long as the convergence can be achieved.

We first drop the term $\partial\left(\boldsymbol{\Phi}_{\boldsymbol{q}}^{\mathrm{T}}\boldsymbol{\lambda}\right)/\partial\boldsymbol{q}$ in the exact Jacobian matrix (21). The term $\boldsymbol{\Phi}_{\boldsymbol{q}}^{\mathrm{T}}\boldsymbol{\lambda}$ physically means the constraint forces, therefore this term actually denotes the constraint stiffness. For a spherical joint connecting with two nodes $i$ and $j$, the constraint equations can be written as $\boldsymbol{\Phi} = \boldsymbol{x}_i - \boldsymbol{x}_j \in \Re^3$, and the constraint Jacobian is a constant matrix taking the form $\boldsymbol{\Phi}_{\boldsymbol{q}} = \left[\cdots \;\; \boldsymbol{I}_3 \;\; \cdots \;\; -\boldsymbol{I}_3 \;\; \cdots\right]$, where $\boldsymbol{I}_3$ is the 3×3 identity matrix. Therefore, the term $\partial\left(\boldsymbol{\Phi}_{\boldsymbol{q}}^{\mathrm{T}}\boldsymbol{\lambda}\right)/\partial\boldsymbol{q}$ is exactly a zero matrix. Abandoning this term would not introduce any error to the Jacobian matrix. For other kinds of joint, this term is typically a small value that would generally not affect the convergence property.

We next focus on the term $\partial \boldsymbol{f}_{\mathrm{INT}} / \partial \boldsymbol{q}$, which is a block diagonal matrix. For each block $\partial \boldsymbol{f}_{\mathrm{INT}}^{(i)} / \partial \boldsymbol{x}_{\mathrm{b}}^{(i)}$, by using expression (14), we have

$$\frac{\partial \boldsymbol{f}_{\mathrm{INT}}^{(i)}}{\partial \boldsymbol{x}_{\mathrm{b}}^{(i)}} = \mathrm{diag}\left(\boldsymbol{R}_{\mathrm{b}}^{(i)}\right) \bar{\boldsymbol{K}}_{\mathrm{b}}^{(i)} \mathrm{diag}\left(\boldsymbol{R}_{\mathrm{b}}^{(i)}\right)^{\mathrm{T}} + \left. \frac{\partial\left( \mathrm{diag}\left(\boldsymbol{R}_{\mathrm{b}}^{(i)}\right) \bar{\boldsymbol{K}}_{\mathrm{b}}^{(i)} \left( \mathrm{diag}\left(\boldsymbol{R}_{\mathrm{b}}^{(i)}\right)^{\mathrm{T}} \boldsymbol{V} - \bar{\boldsymbol{x}}_{\mathrm{b0}}^{(i)} \right)\right)}{\partial \boldsymbol{x}_{\mathrm{b}}^{(i)}} \right|_{\boldsymbol{V} = \boldsymbol{x}_{\mathrm{b}}^{(i)}} \tag{22}$$

where the first term denotes the ordinary orthogonal transformation from the local stiffness matrix and the global one. The second term of expression (22) stems from the variation of the orthogonal matrix to the nodal coordinates of the body. If geometry-based rules such as the three-node side alignment approach (expression (4)) is adopted to determine the orthogonal matrix, the orthogonal matrix is a function of the coordinates of only three nodes, and is independent of the coordinates of all the other nodes belonging to the body. As a result, this term possesses a highly sparse property for a flexible body. This term is also dropped to achieve an effective numerical implementation.

Based on the above operations the Jacobian matrix can now be modified as (the time step index $n$+1 is omitted for brevity)

$$\boldsymbol{J} = \begin{bmatrix} \boldsymbol{C} & \boldsymbol{\Phi}_{\boldsymbol{q}}^{\mathrm{T}} \\ \boldsymbol{\Phi}_{\boldsymbol{q}} & \boldsymbol{0} \end{bmatrix} \tag{23}$$

where $\boldsymbol{C}$ is a large-scale block diagonal matrix given as

$$\boldsymbol{C} = \begin{bmatrix} \boldsymbol{C}^{(1)} & & & \\ & \boldsymbol{C}^{(2)} & & \\ & & \ddots & \\ & & & \boldsymbol{C}^{(N_{\mathrm{b}})} \end{bmatrix} \tag{24}$$

with.

$$\boldsymbol{C}^{(i)} = \bar{\boldsymbol{M}}_{\mathrm{b}}^{(i)} + \alpha h^2 \mathrm{diag}\left(\boldsymbol{R}_{\mathrm{b}}^{(i)}\right) \bar{\boldsymbol{K}}_{\mathrm{b}}^{(i)} \mathrm{diag}\left(\boldsymbol{R}_{\mathrm{b}}^{(i)}\right)^{\mathrm{T}} \tag{25}$$

Since the mass matrix $\bar{\boldsymbol{M}}_{\mathrm{b}}^{(i)}$ is a positive symmetric definite matrix and the local stiffness matrix $\bar{\boldsymbol{K}}_{\mathrm{b}}^{(i)}$ is a positive semi-definite symmetric matrix, each term of $\boldsymbol{C}^{(i)}$ , as well as the entire matrix $\boldsymbol{C}$, is a positive symmetric definite matrix.

## 4.2 Solving the linear system with Schur complement

Due to the including of the orthogonal matrix $\boldsymbol{R}_{\mathrm{b}}^{(i)}$, each term of $\boldsymbol{C}^{(i)}$ is coordinate-dependent. If we directly solve the linear system (20), simply with the newly modified Jacobian matrix, the involved computing effort is still large. To conduct an effective numerical implementation, the second step needed to do is employing the Schur complement to solve the linear system (20). Firstly, the system can be expanded as

$$\begin{cases} \boldsymbol{C}\Delta\ddot{\boldsymbol{q}} + \boldsymbol{\Phi}_{\boldsymbol{q}}^{\mathrm{T}}\Delta\boldsymbol{\lambda} = \boldsymbol{F} \\ \boldsymbol{\Phi}_{\boldsymbol{q}}\Delta\ddot{\boldsymbol{q}} = \boldsymbol{\Psi} \end{cases} \tag{26}$$

By using the first equation and considering that the block $C$ is invertible, we have

$$\Delta\ddot{\boldsymbol{q}} = \boldsymbol{C}^{-1}\boldsymbol{F} - \boldsymbol{C}^{-1}\boldsymbol{\Phi}_{\boldsymbol{q}}^{\mathrm{T}}\Delta\boldsymbol{\lambda} \tag{27}$$

Substituting expression (27) into the second equation of (26),we can first obtain the solution of $\Delta\boldsymbol{\lambda}$

$$\Delta\boldsymbol{\lambda} = \left(\boldsymbol{\Phi}_{\boldsymbol{q}}\boldsymbol{C}^{-1}\boldsymbol{\Phi}_{\boldsymbol{q}}^{\mathrm{T}}\right)^{-1}\left(\boldsymbol{\Phi}_{\boldsymbol{q}}\boldsymbol{C}^{-1}\boldsymbol{F} - \boldsymbol{\Psi}\right) \tag{28}$$

Substituting the above expression back into equation (27), we have the final result of $\Delta\ddot{\boldsymbol{q}}$. The term $\boldsymbol{\Phi}_{\boldsymbol{q}}\boldsymbol{C}^{-1}\boldsymbol{\Phi}_{\boldsymbol{q}}^{\mathrm{T}}$ is the Schur complement of the block $\boldsymbol{C}$ of the matrix $\boldsymbol{J}$. Here, it is assumed that the constraint Jacobian $\boldsymbol{\Phi}_{\boldsymbol{q}}$ is full rank (a condition that is always satisfied for normal multibody dynamic analysis without constraint redundancy), so that the term $\boldsymbol{\Phi}_{\boldsymbol{q}}\boldsymbol{C}^{-1}\boldsymbol{\Phi}_{\boldsymbol{q}}^{\mathrm{T}}$ is invertible. We assume that the number of the constraint equation is much less than the nodal degrees of freedoms, which is usually the case for flexible multibody systems. As a result, the term $\boldsymbol{\Phi}_{\boldsymbol{q}}\boldsymbol{C}^{-1}\boldsymbol{\Phi}_{\boldsymbol{q}}^{\mathrm{T}}$ is a small dimensional matrix, and once it is obtained the inverse matrix (or the decomposition) can be efficiently computed. The problem therefore turns into how to calculate the two terms, $\boldsymbol{C}^{-1}\boldsymbol{F}$ and $\boldsymbol{\Phi}_{\boldsymbol{q}}\boldsymbol{C}^{-1}\boldsymbol{\Phi}_{\boldsymbol{q}}^{\mathrm{T}}$, efficiently.

### 4.3 Effective calculation of $\boldsymbol{C}^{-1}\boldsymbol{F}$

We first focus on the term $\boldsymbol{C}^{-1}\boldsymbol{F}$. Since the large-scale matrix $C$ has a block diagonal structure, it is straightforward that a parallel mode can be used to find the solution,

$$\boldsymbol{C}^{-1}\boldsymbol{F} = \begin{bmatrix} \left(\boldsymbol{C}^{(1)}\right)^{-1}\boldsymbol{F}^{(1)} \\ \left(\boldsymbol{C}^{(2)}\right)^{-1}\boldsymbol{F}^{(2)} \\ \vdots \\ \left(\boldsymbol{C}^{(N_{\mathrm{b}})}\right)^{-1}\boldsymbol{F}^{(N_{\mathrm{b}})} \end{bmatrix} \tag{29}$$

where $\boldsymbol{F}^{(i)}\left(1 \le i \le N_{\mathrm{b}}\right)$ is the force residual error relevant to the $i$th flexible body. However, since each block $\boldsymbol{C}^{(i)}$ is coordinate dependent, if we directly employ expression (25) to computing each component $\left(\boldsymbol{C}^{(i)}\right)^{-1}\boldsymbol{F}^{(i)}$ in (29), the block matrix $\boldsymbol{C}^{(i)}$ should be updated and decomposed at each iterative step, which is still time-consuming.

To fundamentally overcome this problem, we explored the special feature of the block matrix, i.e, the rotational invariance property of the mass matrix, $\mathrm{diag}\left(\boldsymbol{R}_{\mathrm{b}}^{(i)}\right)^{\mathrm{T}}\bar{\boldsymbol{M}}_{\mathrm{b}}^{(i)}\boldsymbol{R}_{\mathrm{b}}^{(i)}\mathrm{diag}\left(\boldsymbol{R}_{\mathrm{b}}^{(i)}\right) \equiv \bar{\boldsymbol{M}}_{\mathrm{b}}^{(i)}$. As a result, we have an extremely important relationship

$$\boldsymbol{C}^{(i)} = \mathrm{diag}\left(\boldsymbol{R}_{\mathrm{b}}^{(i)}\right)\bar{\boldsymbol{C}}^{(i)}\mathrm{diag}\left(\boldsymbol{R}_{\mathrm{b}}^{(i)}\right)^{\mathrm{T}} \tag{30}$$

where matrix $\bar{\boldsymbol{C}}^{(i)}$ is defined as

$$\bar{\boldsymbol{C}}^{(i)} = \bar{\boldsymbol{M}}_{\mathrm{b}}^{(i)} + \alpha h^2 \bar{\boldsymbol{K}}_{\mathrm{b}}^{(i)} \tag{31}$$

Matrix $\bar{\boldsymbol{C}}^{(i)}$ is a constant matrix throughout the analysis, and it is actually the dynamic stiffness matrix term raised in ordinary linear structural dynamic analysis. Besides, it is a sparse positive symmetric definite matrix, and Cholesky decomposition can be conducted

$$\bar{\boldsymbol{C}}^{(i)} = \bar{\boldsymbol{L}}^{(i)}\bar{\boldsymbol{L}}^{(i)\mathrm{T}} \tag{32}$$

where $\bar{\boldsymbol{L}}^{(i)}$ is a lower triangular matrix. Matrices $\bar{\boldsymbol{C}}^{(i)}$ and $\bar{\boldsymbol{L}}^{(i)}$ are typically sparse matrix. To further enhance the computational efficiency, it is useful to improve the sparsity of the triangular matrix $\bar{\boldsymbol{L}}^{(i)}$, which can be achieved by reordering matrix $\bar{\boldsymbol{C}}^{(i)}$ before making the decomposition. Some popular algorithms have been proposed, such as the nested dissection permutation (dissect) (Brainman and Toledo, 2002), approximate minimum degree permutation (amd) (Davis et al., 2004) and sparse reverse Cuthill-McKee ordering (symrcm) (Cuthill and McKee, 1969) methods. Based on our tests, it appears that the dissect algorithm typically incurs the least amount of fill-in of nonzero entities, at least for the numerical examples to be tested in this paper. This ordering method is adopted for all the numerical examples presented in Section 6.

The inverse matrix of $\boldsymbol{C}^{(i)}$ can now be formally written as

$$\left(\boldsymbol{C}^{(i)}\right)^{-1} = \mathrm{diag}\left(\boldsymbol{R}_{\mathrm{b}}^{(i)}\right)\left(\bar{\boldsymbol{L}}^{(i)-1}\right)^{\mathrm{T}}\left(\bar{\boldsymbol{L}}^{(i)-1}\right)\mathrm{diag}\left(\boldsymbol{R}_{\mathrm{b}}^{(i)}\right)^{\mathrm{T}} \tag{33}$$

and the term $\left(\boldsymbol{C}^{(i)}\right)^{-1}\boldsymbol{F}^{(i)}$ can be obtained following four sequential steps

$$\left(\boldsymbol{C}^{(i)}\right)^{-1}\boldsymbol{F}^{(i)}=\underbrace{\mathrm{diag}\left(\boldsymbol{R}_{\mathrm{b}}^{(i)}\right)\left(\underbrace{\left(\bar{\boldsymbol{L}}^{(i)-1}\right)^{\mathrm{T}}\left(\underbrace{\left(\bar{\boldsymbol{L}}^{(i)-1}\right)\left(\underbrace{\mathrm{diag}\left(\boldsymbol{R}_{\mathrm{b}}^{(i)}\right)^{\mathrm{T}}\boldsymbol{F}^{(i)}}_{1}\right)}_{2}\right)}_{3}\right)}_{4} \tag{34}$$

The above expression involves two kinds of basic operation that is worth mentioning. The first one is the multiplication of a block diagonal orthogonal matrix with a vector (steps 1 and 4), e.g. diag($\boldsymbol{R}$)$\boldsymbol{V}$. To accelerate the calculation and save memory, in real numerical implementation there is no need to rigidly first formulate the large-scale block diagonal orthogonal matrix and then make the multiplication. An effective operation is through reshaping: assuming that $\boldsymbol{V}\in\Re^{3m\times1}$, we can first reshape the vector into a matrix $\boldsymbol{V}_{\mathrm{M}}\in\Re^{3\times m}$, where each column representing the term related to a particular node , then make the multiplication $\left(\boldsymbol{R}\boldsymbol{V}_{\mathrm{M}}\right)\in\Re^{3\times m}$, and finally reshape the outcome back to a $3m\times1$ vector. The second kind of basic operation is the multiplication of the inverse of the triangular matrix with a vector (steps 2 and 3), e.g. $\bar{\boldsymbol{L}}^{-1}\boldsymbol{V}$ . This formulation only has a mathematical meaning. In numerical implementation this is no need to first obtain the large-scale inverse matrix and then make the multiplication, and the result is effectively obtained through back substitution.

## 4.4 Effective calculation of $\boldsymbol{\Phi}_{\boldsymbol{q}}\boldsymbol{C}^{-1}\boldsymbol{\Phi}_{\boldsymbol{q}}^{\mathrm{T}}$

We next focus on calculating the term $\boldsymbol{\Phi}_{\boldsymbol{q}}\boldsymbol{C}^{-1}\boldsymbol{\Phi}_{\boldsymbol{q}}^{\mathrm{T}}$. Intuitively, one may following the previous method: by considering each column of $\boldsymbol{\Phi}_{\boldsymbol{q}}^{\mathrm{T}}$ as a $\boldsymbol{F}$ vector in the term $\boldsymbol{C}^{-1}\boldsymbol{\Phi}_{\boldsymbol{q}}^{\mathrm{T}}$ can be first obtained, then make the multiplication with $\boldsymbol{\Phi}_{\boldsymbol{q}}$. However, to make the calculation even faster, we can explore the nature of this term.

With Eq.(33) and considering the block diagonal structure, the inverse of matrix of $\boldsymbol{C}$ can be written as

$$\left(\boldsymbol{C}\right)^{-1}=\boldsymbol{D}\left(\bar{\boldsymbol{L}}^{-1}\right)^{\mathrm{T}}\left(\bar{\boldsymbol{L}}^{-1}\right)\boldsymbol{D}^{\mathrm{T}} \tag{35}$$

where $\bar{\boldsymbol{L}}$ and $\boldsymbol{D}$ are both block diagonal matrices given as

$$\bar{\boldsymbol{L}}^{-1} = \begin{bmatrix} \left(\bar{\boldsymbol{L}}^{(1)}\right)^{-1} & & \\ & \ddots & \\ & & \left(\bar{\boldsymbol{L}}^{(N_\mathrm{b})}\right)^{-1} \end{bmatrix} \tag{36}$$

$$\boldsymbol{D} = \begin{bmatrix} \mathrm{diag}\left(\boldsymbol{R}_\mathrm{b}^{(1)}\right) & & \\ & \ddots & \\ & & \mathrm{diag}\left(\boldsymbol{R}_\mathrm{b}^{(N_\mathrm{b})}\right) \end{bmatrix} \tag{37}$$

with Eq. (35), we have

$$\boldsymbol{\Phi}_{\boldsymbol{q}}\boldsymbol{C}^{-1}\boldsymbol{\Phi}_{\boldsymbol{q}}^\mathrm{T} = \boldsymbol{\Phi}_{\boldsymbol{q}}\boldsymbol{D}\left(\bar{\boldsymbol{L}}^{-1}\right)^\mathrm{T}\left(\bar{\boldsymbol{L}}^{-1}\right)\boldsymbol{D}^\mathrm{T}\boldsymbol{\Phi}_{\boldsymbol{q}}^\mathrm{T} = \left(\bar{\boldsymbol{L}}^{-1}\boldsymbol{D}^\mathrm{T}\boldsymbol{\Phi}_{\boldsymbol{q}}^\mathrm{T}\right)^\mathrm{T}\left(\bar{\boldsymbol{L}}^{-1}\boldsymbol{D}^\mathrm{T}\boldsymbol{\Phi}_{\boldsymbol{q}}^\mathrm{T}\right) \tag{38}$$

Therefore, one only need to compute the term $\bar{\boldsymbol{L}}^{-1}\boldsymbol{D}^\mathrm{T}\boldsymbol{\Phi}_{\boldsymbol{q}}^\mathrm{T}$ and make the multiplication with the transpose of itself. By acting in this way, the operation of one back substitution is saved. The term $\bar{\boldsymbol{L}}^{-1}\boldsymbol{D}^\mathrm{T}\boldsymbol{\Phi}_{\boldsymbol{q}}^\mathrm{T}$ can be obtained by first calculating $\boldsymbol{D}^\mathrm{T}\boldsymbol{\Phi}_{\boldsymbol{q}}^\mathrm{T}$ and then make the back substitution. However, for some cases, this kind of operation is still not that effective. For example, if the system consisting of a relatively large number of spherical joint, the corresponding $\boldsymbol{\Phi}_{\boldsymbol{q}}^\mathrm{T}$ is a large-scale constant matrix (this situation occurs for a flexible body that is subjected to large deformation, and one body one component may not able to accurately analyze the body. In this case, one can divide the body into several components, and the interfaces are modeled using considerable spherical joints). Considering that both $\boldsymbol{\Phi}_{\boldsymbol{q}}^\mathrm{T}$ and $\bar{\boldsymbol{L}}^{-1}$ are constant matrices, the result of $\bar{\boldsymbol{L}}^{-1}\boldsymbol{D}^\mathrm{T}\boldsymbol{\Phi}_{\boldsymbol{q}}^\mathrm{T}$ is purely a function of matrix $\boldsymbol{D}$. Suppose that the sub block $\boldsymbol{R}_\mathrm{b}^{(i)} \in \Re^{3\times3}$ in (37) takes the form

$$\boldsymbol{R}_\mathrm{b}^{(i)} = \begin{bmatrix} r_{1,1}^{(i)} & r_{1,2}^{(i)} & r_{1,3}^{(i)} \\ r_{2,1}^{(i)} & r_{2,2}^{(i)} & r_{2,3}^{(i)} \\ r_{3,1}^{(i)} & r_{3,2}^{(i)} & r_{3,3}^{(i)} \end{bmatrix} \tag{39}$$

According to the expression of (37), matrix $\boldsymbol{D}$ consists of $9N_\mathrm{b}$ different entities, and can be decomposed as

$$\boldsymbol{D} = \sum_{1\le k\le N_\mathrm{b}}\sum_{1\le j\le 3}\sum_{1\le i\le 3}\boldsymbol{D}_{i,j}^{(k)} r_{i,j}^{(k)} \tag{40}$$

where $\boldsymbol{D}_{i,j}^{(k)}$ is a 0-1 matrix that is constructed similar to $\boldsymbol{D}$, but replacing all $r_{i,j}^{(k)}$

terms with unities and all other terms with zeros. As a result, one can precomputed the constant terms $\bar{\boldsymbol{L}}^{-1}\boldsymbol{D}_{i,j}^{(k)\mathrm{T}}\boldsymbol{\Phi}_{\boldsymbol{q}}^{\mathrm{T}}$ once, and make linear combinations with the real time $r_{i,j}^{(k)}$.

$$\bar{\boldsymbol{L}}^{-1}\boldsymbol{D}^{\mathrm{T}}\boldsymbol{\Phi}_{\boldsymbol{q}}^{\mathrm{T}} = \sum_{1\le k\le N_{\mathrm{b}}}\sum_{1\le j\le 3}\sum_{1\le i\le 3}\left(\bar{\boldsymbol{L}}^{-1}\boldsymbol{D}_{i,j}^{(k)\mathrm{T}}\boldsymbol{\Phi}_{\boldsymbol{q}}^{\mathrm{T}}\right)r_{i,j}^{(k)} \tag{41}$$

Actually, the term $\bar{\boldsymbol{L}}^{-1}\boldsymbol{D}_{i,j}^{(k)\mathrm{T}}\boldsymbol{\Phi}_{\boldsymbol{q}}^{\mathrm{T}}$ possesses a very sparsity property. The about statement merely explains the basic mathematical principle for seeking a faster calculation. In real numerical implementation, one may use the assembly method to obtain the result, that is, by meticulously investigating and recoding the non-zero blocks of each term, and assuming the blocks according to the essence of Eq. (41).

For other kinds of joints, $\boldsymbol{\Phi}_{\boldsymbol{q}}^{\mathrm{T}}$ is typically coordinate-dependent, but is still highly sparse. Similar method can be conducted by considering $\boldsymbol{D}^{\mathrm{T}}\boldsymbol{\Phi}_{\boldsymbol{q}}^{\mathrm{T}}$ as an ensemble, and exploring its linear combination with the non-zero entities, i.e., find $\boldsymbol{D}^{\mathrm{T}}\boldsymbol{\Phi}_{\boldsymbol{q}}^{\mathrm{T}} = \sum \kappa_j \boldsymbol{\Theta}_j$, where $\boldsymbol{\Theta}_j$ represents a series of sparse constant matrix. By precomputing each constant matrix $\bar{\boldsymbol{L}}^{-1}\boldsymbol{\Theta}_j$ and make the linear combination with the time-varying coefficients $\kappa_j$, the final result can be obtained as $\bar{\boldsymbol{L}}^{-1}\boldsymbol{D}^{\mathrm{T}}\boldsymbol{\Phi}_{\boldsymbol{q}}^{\mathrm{T}} = \sum \kappa_j\left(\bar{\boldsymbol{L}}^{-1}\boldsymbol{\Theta}_j\right)$.

By using the above operations, the term $\boldsymbol{\Phi}_{\boldsymbol{q}}\boldsymbol{C}^{-1}\boldsymbol{\Phi}_{\boldsymbol{q}}^{\mathrm{T}}$ can be calculated effectively. However, considering that implicit time integration is used, this term should be updated at every iteration, which may devalue the efficiency of the time integration. A useful tip for further accelerating the time integration algorithm is that one may only update the term $\boldsymbol{\Phi}_{\boldsymbol{q}}\boldsymbol{C}^{-1}\boldsymbol{\Phi}_{\boldsymbol{q}}^{\mathrm{T}}$ (or its inverse matrix/or its decomposition) at every time step, instead of at every iteration. According to the authors' numerical tests, this would almost not influence on the convergence performances for the numerical examples tested in this paper.

## 5. Some discussions related to the proposed framework

### 5.1 Simplification of the Jacobian matrix

The above section illustrates how to conduct the effective numerical implementation for the proposed framework. By using the proposed implementations, the necessity of ceaselessly forming and decomposing the time-varying Jacobian matrix in conventional flexible body analysis is ultimately avoided. Therefore, it is expected that the proposed framework is able to achieve an extremely efficiency, even with a relatively large number of system DOFs. The premise of doing so lies that we exploit the special nature of the CR modeling method.

It can be found that the simplification of the Jacobian matrix possesses a foundational step to the proposed numerical implementation. One may wonder that is it possible to further simplify the Jacobian matrix by completely dropping the term $\partial \boldsymbol{f}_{\mathrm{INT}}/\partial \boldsymbol{q}$ and only the mass matrix is retained (i.e., $\boldsymbol{C}=\boldsymbol{M}$ in (23)), and thus the solution of the linear system (20) might be even faster to be solved. Unfortunately, this is only feasible for the analysis with an extremely small time step size. For a relatively large time step size, convergence generally cannot be achieved with this excessive simplified Jacobian matrix. A mathematical interpretation of this phenomenon is that according to Eq. (25) matrix $\boldsymbol{C}$ get close to $\boldsymbol{M}$ as $h \to 0$. The limitation of using extremely small time step size however devalues the original intention of using implicit algorithms.

### 5.2 Damping force and static analysis

All the proceeding discussions focus on dynamic analysis. Static analysis (static equilibrium problem) may also be concerned in multibody analysis. Conventional static equilibrium analysis for multibody system can be conducted by directly focusing on the equilibrium equation, which is a set of nonlinear equations that can be obtained by dropping all the dynamic terms in the dynamic governing equation (16). The solutions therefore can be obtained by classical Newton-Raphson method with appropriate choice of the initial guess. For the proposed framework, the modeling method presented in Section 3 is still feasible, however, the effective numerical implementations presented in Section 4 is not available, which is due to the following fact. The absence of mass matrix in static analysis leads the upper left block matrix $\boldsymbol{C}$ in the total Jacobin matrix being purely the tangent stiffness matrix. Since no boundary condition is applied to the component to prevent its overall rigid body motions, matrix $\boldsymbol{C}$ is singular. To overcome this problem, one may employ the dynamic relaxation approach for static analysis (Barnes, 1999; Brew and Brotton, 1971). This is done by introducing some pseudo damping forces to the system and applying the dynamic analysis. The system finally evolves to its static equilibrium state due to the energy dissipation of those damping forces. This kind of dynamic relaxation approach typically requires more iterations than the classical Newton-Raphson-based method. However, due to the effective numerical implementations presented in Section 4, it is expected that the solution can be still be very effectively obtained.

Two kinds of damping force are applicable, the first one is simply let the generalized damping of $i$th body be

$$\boldsymbol{f}_{\mathrm{Dam}}^{(i)}\left(\boldsymbol{x}_{\mathrm{b}}^{(i)}, \dot{\boldsymbol{x}}_{\mathrm{b}}^{(i)}\right)=\zeta \dot{\boldsymbol{x}}_{\mathrm{b}}^{(i)} \tag{42}$$

where $\zeta(>0)$ is an appropriate damping coefficient. The second one is to follow a CR manner given as

$$\boldsymbol{f}_{\mathrm{Dam}}^{(i)}\left(\boldsymbol{x}_{\mathrm{b}}^{(i)},\dot{\boldsymbol{x}}_{\mathrm{b}}^{(i)}\right)=\mathrm{diag}\left(\boldsymbol{R}_{\mathrm{b}}^{(i)}\right)\bar{\boldsymbol{D}}_{\mathrm{b}}^{(i)}\mathrm{diag}\left(\boldsymbol{R}_{\mathrm{b}}^{(i)}\right)^{\mathrm{T}}\dot{\boldsymbol{x}}_{\mathrm{b}}^{(i)} \tag{43}$$

where $\bar{\boldsymbol{D}}_{\mathrm{b}}^{(i)}=a\bar{\boldsymbol{M}}_{\mathrm{b}}^{(i)}+b\bar{\boldsymbol{K}}_{\mathrm{b}}^{(i)}$ is the conventional Rayleigh proportional damping matrix of the body ($a$ and $b$ are positive scale coefficient). The first kind of damping force is computationally more effective and is more suitable for pure dynamic relaxation analysis, while the second one is more suitable for modeling flexible multibody system with real structural damping. Since the damping force is a function of the generalized velocity, the Jacobian matrix (23) should be modified by taking into account the contribution of this kind of force. In context of the current Newmark algorithm, the sub-block $\bar{\boldsymbol{C}}^{(i)}$ originally given through Eq.(31) should be replaced respectively as

$$\bar{\boldsymbol{C}}^{(i)}=\bar{\boldsymbol{M}}_{\mathrm{b}}^{(i)}+\zeta\delta h\boldsymbol{I}_{\mathrm{b}}^{(i)}+\alpha h^{2}\bar{\boldsymbol{K}}_{\mathrm{b}}^{(i)} \tag{44}$$

$$\bar{\boldsymbol{C}}^{(i)}=\bar{\boldsymbol{M}}_{\mathrm{b}}^{(i)}+\delta h\bar{\boldsymbol{D}}_{\mathrm{b}}^{(i)}+\alpha h^{2}\bar{\boldsymbol{K}}_{\mathrm{b}}^{(i)} \tag{45}$$

In (44) $\boldsymbol{I}_{\mathrm{b}}^{(i)}$ is an identity matrix that is of the same dimension as the mass matrix.

### 5.3 Coordinate-dependent external force

In the proceeding sections, we only consider coordinate-independent external forces in the dynamic analysis. Commonly encountered external forces such as gravity force, spatially-fixed applied force belong to this kind of force. For these forces, there is no contribution to the system stiffness matrix (Jacobian matrix) in the analysis. In some cases, coordinate-dependent external force may also be involved in multibody system, for example, the applied body-fixed force (e.g. a force applied on a node and the direction always point to another node), the contact forces between the body and the environment, or between two bodies. In these cases, the Jacobian term (external force stiffness matrix) $\partial\boldsymbol{f}_{\mathrm{EXT}}/\partial\boldsymbol{q}$ is not exactly a zero matrix, and should be discussed. For the coordinate-dependent external force, if the force magnitude is relatively small compared with that of the internal force. The stiffness matrix contributed by the force can be directly neglected which would not influence the convergence. Otherwise, the external force stiffness matrix should be taken into account. Two approaches are provided.

The first approach addresses the problem from algorithm point of view. Suppose that the force is applied on a particulate node at the $i$th body and the force only depends of the nodes belongs to the body. The external force stiffness matrix is also coordinate-dependent and can be intentionally written as a form $\boldsymbol{K}_{\mathrm{EXT}}^{(i)}=\mathrm{diag}\left(\boldsymbol{R}_{\mathrm{b}}^{(i)}\right)\underbrace{\mathrm{diag}\left(\boldsymbol{R}_{\mathrm{b}}^{(i)}\right)^{\mathrm{T}}\boldsymbol{K}_{\mathrm{EXT}}^{(i)}\mathrm{diag}\left(\boldsymbol{R}_{\mathrm{b}}^{(i)}\right)}_{=\bar{\boldsymbol{K}}_{\mathrm{EXT}}^{(i)}}\mathrm{diag}\left(\boldsymbol{R}_{\mathrm{b}}^{(i)}\right)^{\mathrm{T}}$, while $\bar{\boldsymbol{K}}_{\mathrm{EXT}}^{(i)}$ is now a coordinate-dependent matrix. By adding this term into the Jacobian matrix (23), the corresponding block matrix $\bar{\boldsymbol{C}}^{(i)}$ original given through (31) now turns into

$$\bar{\boldsymbol{C}}^{(i)}=\bar{\boldsymbol{M}}_{\mathrm{b}}^{(i)}+\alpha h^{2}\bar{\boldsymbol{K}}_{\mathrm{b}}^{(i)}+\underline{\alpha h^{2}\bar{\boldsymbol{K}}_{\mathrm{EXT}}^{(i)}} \tag{46}$$

The term $\bar{\boldsymbol{C}}^{(i)}$ is no longer a constant matrix due to the additional external force stiffness term $\bar{\boldsymbol{K}}_{\mathrm{EXT}}^{(i)}$, and it should be noted that this term is not necessarily symmetric. Direct (LU) decomposition of such a large-scale matrix at every iteration in dynamic analysis is rather expensive. To accelerate the analysis, one may first note that $\bar{\boldsymbol{K}}_{\mathrm{EXT}}^{(i)}$ is highly sparse and only a limited number of DOFs in $\bar{\boldsymbol{C}}^{(i)}$ are influenced by the external force stiffness, therefore by properly recording the DOFs, the block matrix $\bar{\boldsymbol{C}}^{(i)}$ can be partitioned into

$$\bar{\boldsymbol{C}}^{(i)} = \begin{bmatrix} \bar{\boldsymbol{C}}_{11} & \bar{\boldsymbol{C}}_{12} \\ \bar{\boldsymbol{C}}_{21} & \bar{\boldsymbol{C}}_{22} \end{bmatrix} \tag{47}$$

where $\bar{\boldsymbol{C}}_{11}$ a large-scale constant matrix can be pre-decomposed, while $\bar{\boldsymbol{C}}_{22}$ is a small-scale coordinate-dependent matrix that is related to the DOFs of the applied force, and $\bar{\boldsymbol{C}}_{12}$ and $\bar{\boldsymbol{C}}_{21}$ are highly sparse rectangular matrices. We thus explicitly have

$$\left(\bar{\boldsymbol{C}}^{(i)}\right)^{-1} = \begin{bmatrix} \boldsymbol{I} & -\bar{\boldsymbol{C}}_{11}^{-1}\bar{\boldsymbol{C}}_{12} \\ \boldsymbol{0} & \boldsymbol{I} \end{bmatrix} \begin{bmatrix} \bar{\boldsymbol{C}}_{11}^{-1} & \boldsymbol{0} \\ \boldsymbol{0} & \Delta_{\bar{C}_{11}}^{-1} \end{bmatrix} \begin{bmatrix} \boldsymbol{I} & \boldsymbol{0} \\ -\bar{\boldsymbol{C}}_{21}\bar{\boldsymbol{C}}_{11}^{-1} & \boldsymbol{I} \end{bmatrix} \tag{48}$$

where $\Delta_{\bar{C}_{11}} = \bar{\boldsymbol{C}}_{22} - \bar{\boldsymbol{C}}_{21}\bar{\boldsymbol{C}}_{11}^{-1}\bar{\boldsymbol{C}}_{12}$ is the Schur complement of $\bar{\boldsymbol{C}}_{11}$ in $\bar{\boldsymbol{C}}^{(i)}$ and it is a small-scale square matrix, thus the inverse matrix can be effectively calculated. The terms $\left(\boldsymbol{C}^{(i)}\right)^{-1}\boldsymbol{F}^{(i)}$, and $\boldsymbol{\Phi}_{\boldsymbol{q}}\boldsymbol{C}^{-1}\boldsymbol{\Phi}_{\boldsymbol{q}}^{\mathrm{T}}$ can then be updated with the above expression.

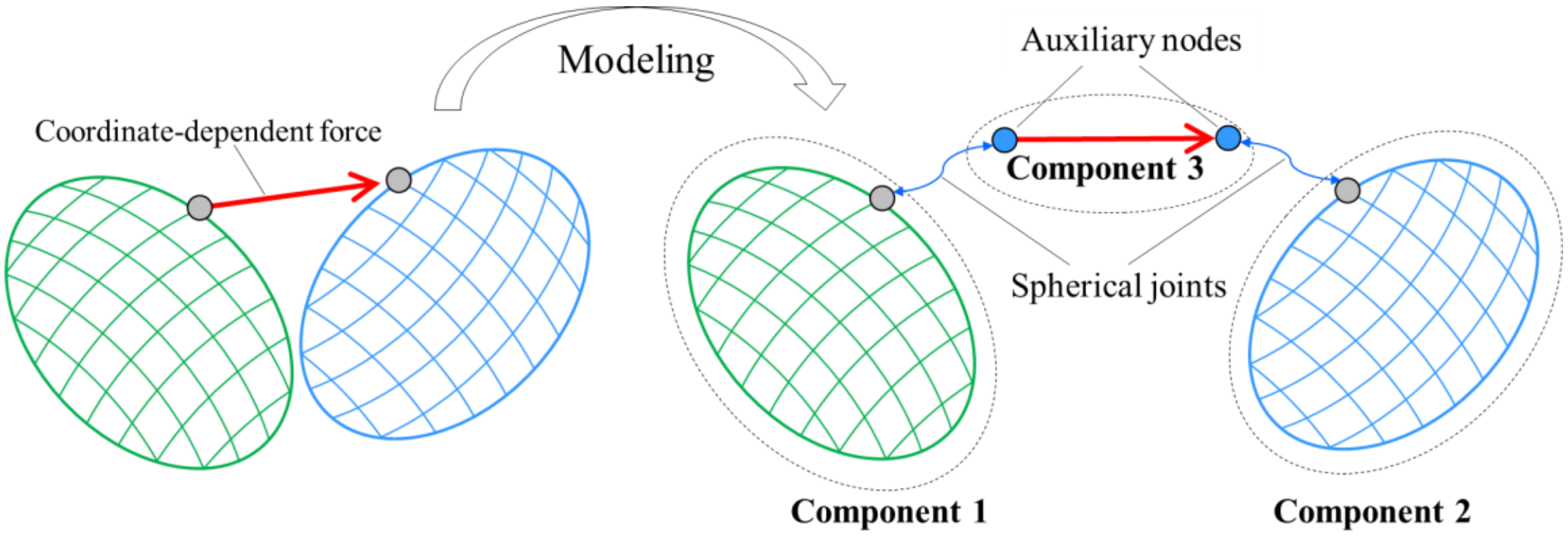

Fig. 3 Modeling of coordinate-dependent force.

The above approach addresses the problem from algorithm point of view, while the issue can also be addressed using another approach, that is, from the modeling phase. As shown in Fig. 3, within this approach some additional/auxiliary nodes are first introduced. These auxiliary nodes connect with the original flexible body through spherical joints. The applied force originally acting on the node of a particular flexible body can now be modeled as acting on these additional nodes. Besides, the direction and magnitude of the force purely depends on these auxiliary nodes. Therefore, the applied force can be considered as a "small scale component" this is isolated with the original flexible body. The only connections between the applied force and the bodies are through the auxiliary nodes and the additional spherical joints. To avoid singularity of the dynamic stiffness some nodal mass belongs to the original flexible body should be assigned to these auxiliary nodes. With these operations, the $\boldsymbol{C}$ matrix appeared in the Jacobin matrix (23) now is augmented into

$$\boldsymbol{C} = \left[\begin{array}{ccc|c} \boldsymbol{C}^{(1)} & & & \\ & \ddots & & \\ & & \boldsymbol{C}^{(N_b)} & \\ \hline & & & \boldsymbol{C}^{(F)} \end{array}\right] \tag{49}$$

where $\boldsymbol{C}^{(F)}$ is dynamic stiffness matrix associated with the coordinates of the auxiliary nodes. Although $\boldsymbol{C}^{(F)}$ is coordinate-dependent, its inverse or decomposition can be effectively evaluated as it is a small scale square matrix. In summary, this kind of approach is more straightforward and agile. The price to pay is the number of system generalized coordinates being slightly increased.

5.5 Further discussions on the proposed framework

From above discussions, it can be found that the proposed framework is especially suitable for those multibody systems with less constraint equations but with a large-scale node coordinates. Each component in the system can be considered as small deformed such that a single CR frame is able to describe the large overall motion of the component. Fortunately, many systems in mechanical engineering field belong to this kind system. For some systems with large deformed components, one may also use the proposed framework by breaking a single large-deformed component into more non-overlap small-deformed components and these non-overlap components are connected through spherical joints on the interfaces, as shown in Fig. 4. The weakness of the proposed framework in this situation is that the number of system constraint equations raises due to the additional spherical joints on the interfaces and the efficiency of the proposed approach is devalued (in the process of solving the Schur complement equation (28)).

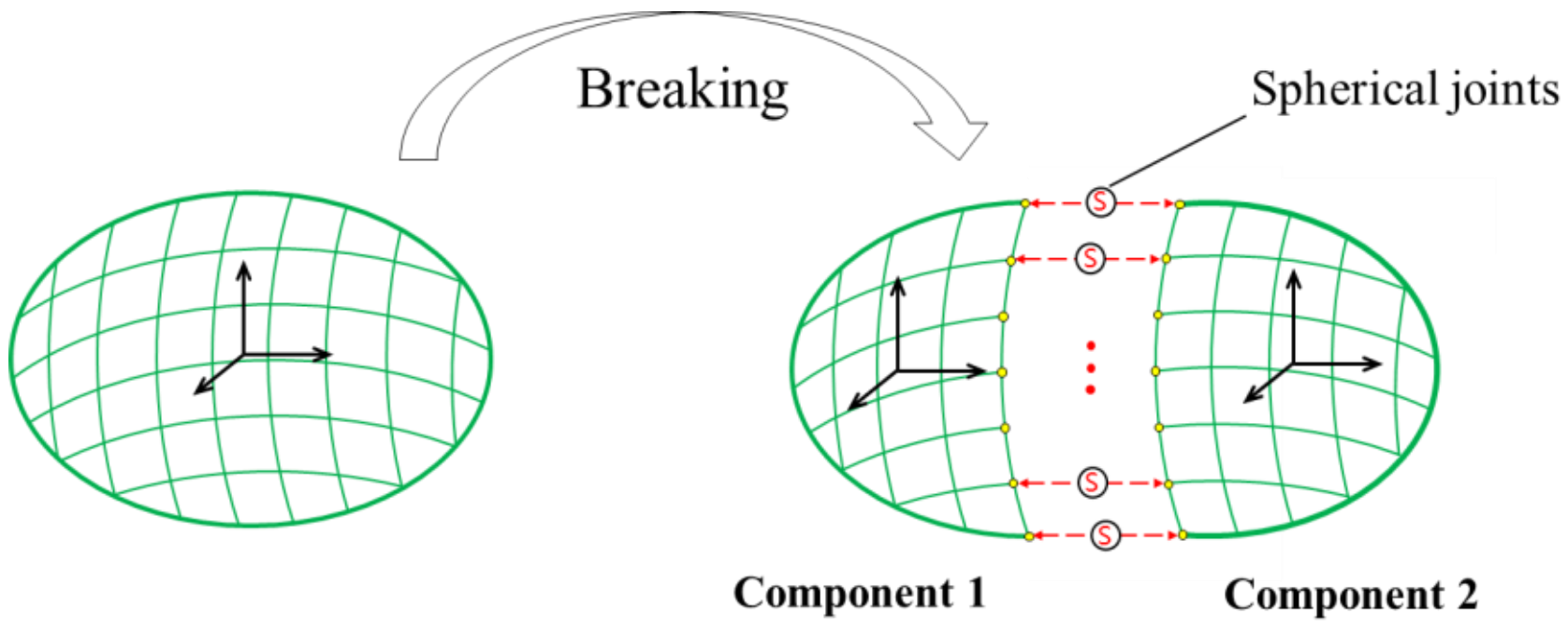

Fig. 4 Breaking a component into more non-overlap small-deformed components.

In limiting case, the proposed framework can degenerate to conventional CR formulation in the modeling case that each single element is considered as a component, and different elements are connected through a large number of spherical joints. In this situation there is no advantage to use the numerical implementation approach presented in section 4, since the price to solve the Schur complement equation (28) exceed the price to solve the system linear equation (26) itself.

## 6. Numerical examples

In this section multiple examples are presented to demonstrate the performance of the proposed framework, in terms of accuracy, efficiency and convergence property. All the examples are carried out in MATLAB 2019b based on the Intel Core i7-1165G7 CPU running on 64-bit operating system at 2.80 GHz.

For all examples, the components are assumed made of aluminum with parameters: elastic modulus $E$ = 70e9; Poisson's ration $\nu$ = 0.3; density $\rho$ = 2700. The International System of Units (m-kg-s) is used. Generally, the results obtained by three different approaches are compared:

(1) "Proposed CR" approach, denoting the nonlinear dynamic analysis using the proposed component-level CR framework.

(2) "Full CR" approach, denoting the conventional nonlinear dynamic analysis where each element is modeled using a common CR element.

(3) "Ansys" approach. To highlight the correctness of the obtained results, the analysis is also carried out with commercial software Ansys (using Solid 45 elements with KEYOPT(1)=1, and NLGEOM=On).

For all examples, the systems are meshed with bilinear eight-node hexahedral elements with full integration. In all dynamic analyses, the Newmark algorithm is adopted with parameters $\alpha$=0.26, $\delta$=0.5. With those parameters, no algorithm damping is introduced. For all the iterations, the convergence criterion is defined as $\|\boldsymbol{G}_{\mathrm{RES}}\| \le \varepsilon$, where $\varepsilon$ denotes the convergence tolerance.

### 6.1 Cuboid pendulum under 3D gravitational field

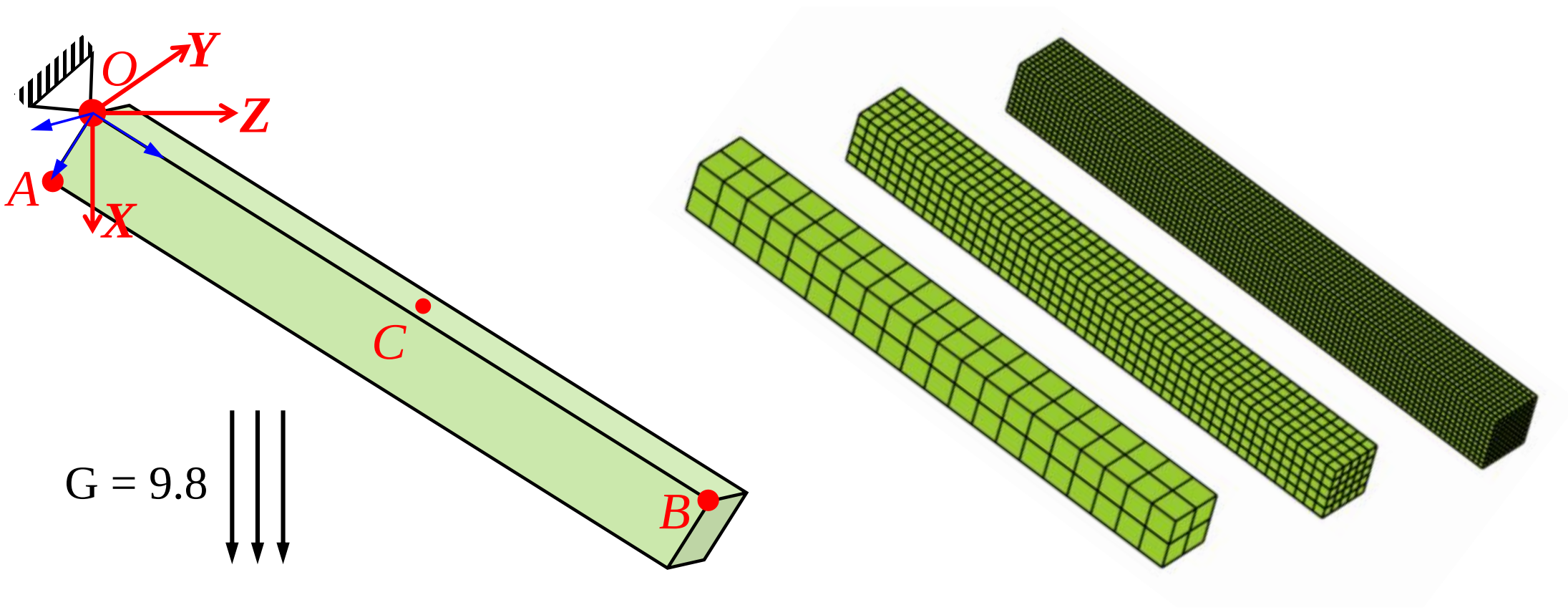

Fig. 5 Cuboid pendulum under 3D gravitational field (left); the three finite element models to be considered (right).

The first example focuses on a cuboid pendulum under 3D gravitational field, as shown in Fig. 5. *OXYZ* denotes the global coordinate system. The pendulum is initially placed along the horizontal direction (z-axis), with one corner being fixed to the space. Due to the effect of gravity, the system undergoes a large overall motion. The geometric parameters of cuboid pendulum is 40×4×4. In order to test the accuracy of the propose framework, the analyses are first conducted using the previously mentioned three methods with a coarse mesh (80 elements, 567 DOFs). For the proposed CR approach, the pendulum is modeled as a single flexible body (component), and the local CR frame is determined using the positions of the three corners (A-O-B) with Eq. (4). All analyses are conducted with a fixed time step size: $h$=0.01. The convergence tolerance of the Newton-Raphson iteration is set as $5\times10^{-3.}$

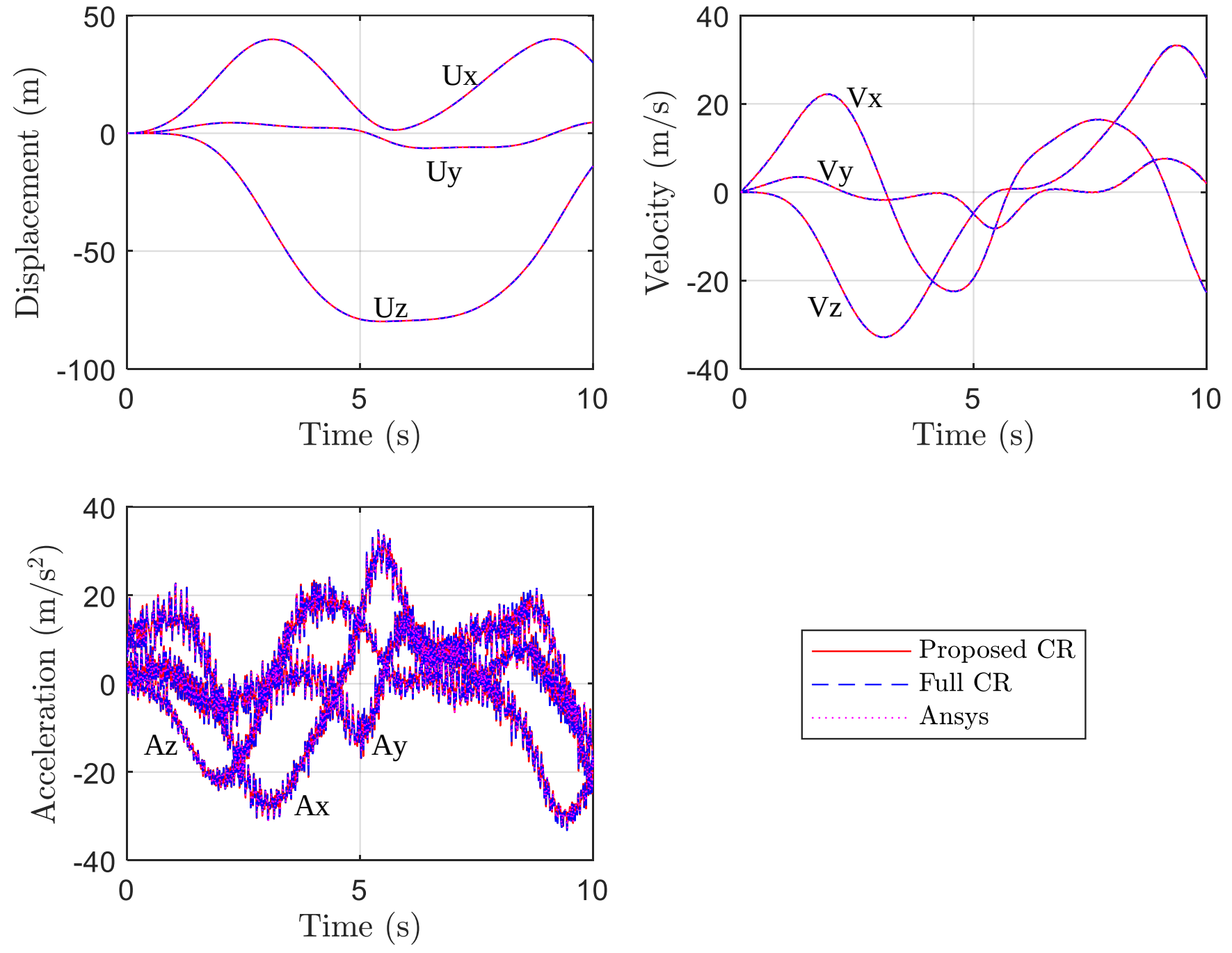

Fig. 6 Time history curves of the global displacements, velocities and accelerations of point B in the *X-Y-Z* directions.

Fig. 6 gives the time history curves of the global displacement, velocity and acceleration of point B in the *X-Y-Z* directions. It is found that the three different methods yield almost the same results. Due to elastic effects, the acceleration curves experience high frequency oscillation coupled with the large overall motion. To conduct a deep comparison from the elastic deformation viewpoint, Fig. 7 gives the local displacements of the mid-span point C in the x-y-z directions. For the "Full CR" and "Ansys" methods, the local displacements are calculated based on the respective global results following the same rule as that for the proposed CR approach. It is found that the local displacements agree well with each other. The displacement is at the level of 1e-3 order of magnitude which implies that the structure is subjected to small deformations during the large overall motion. These results demonstrate the correctness of the proposed framework. To give an insight into the convergence performances of the proposed framework, Fig. 8 presents the number of iterations at each time step. It is found that convergence can be achieved in average 4 iterations at each time step. This outcome shows that even with a relatively large time step size (0.01s), the proposed framework can still ensure a fast and stable convergence performance.

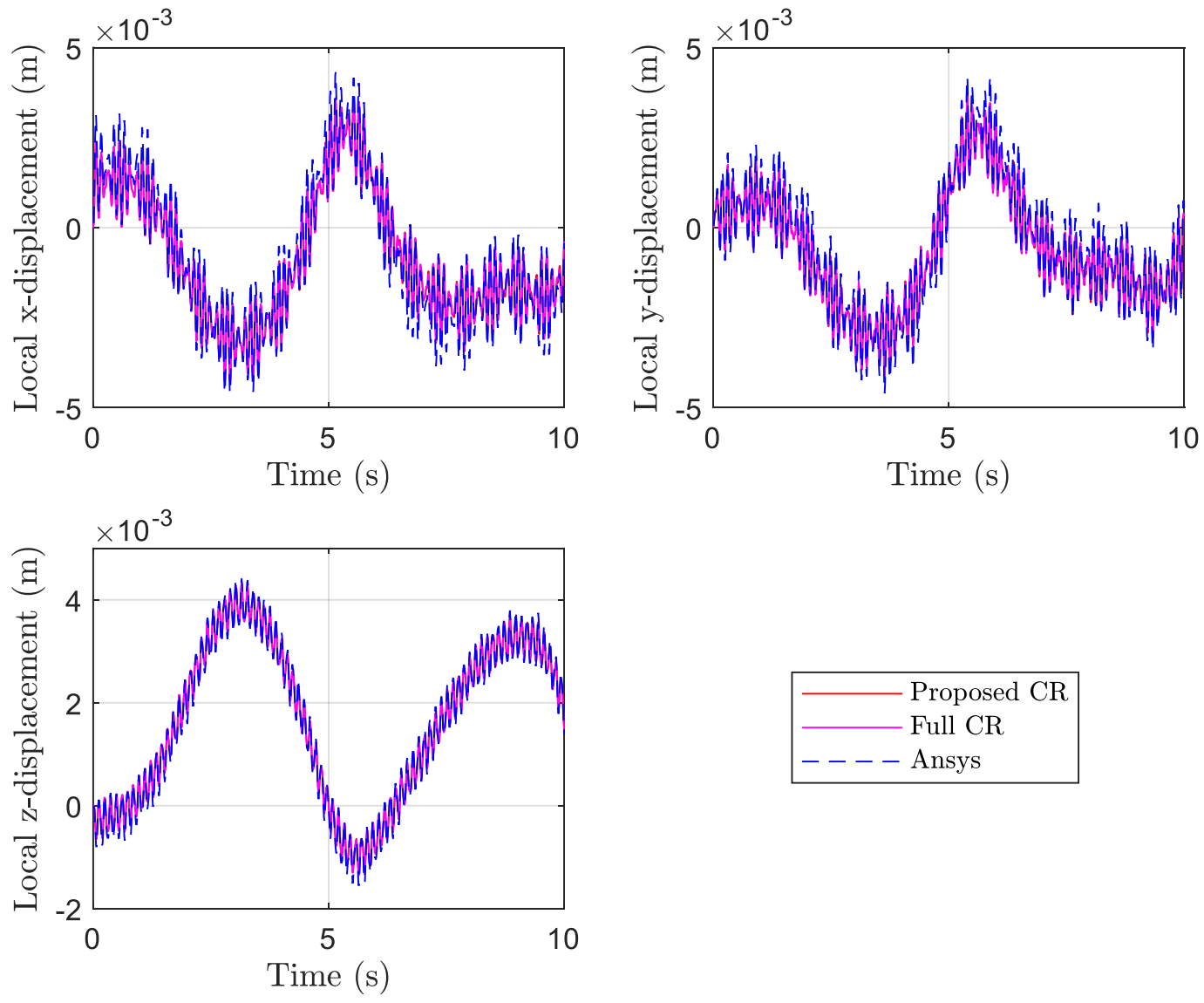

Fig. 7 Time history curves of local displacements of the mid-span point C in the *x*-*y*-*z* directions.

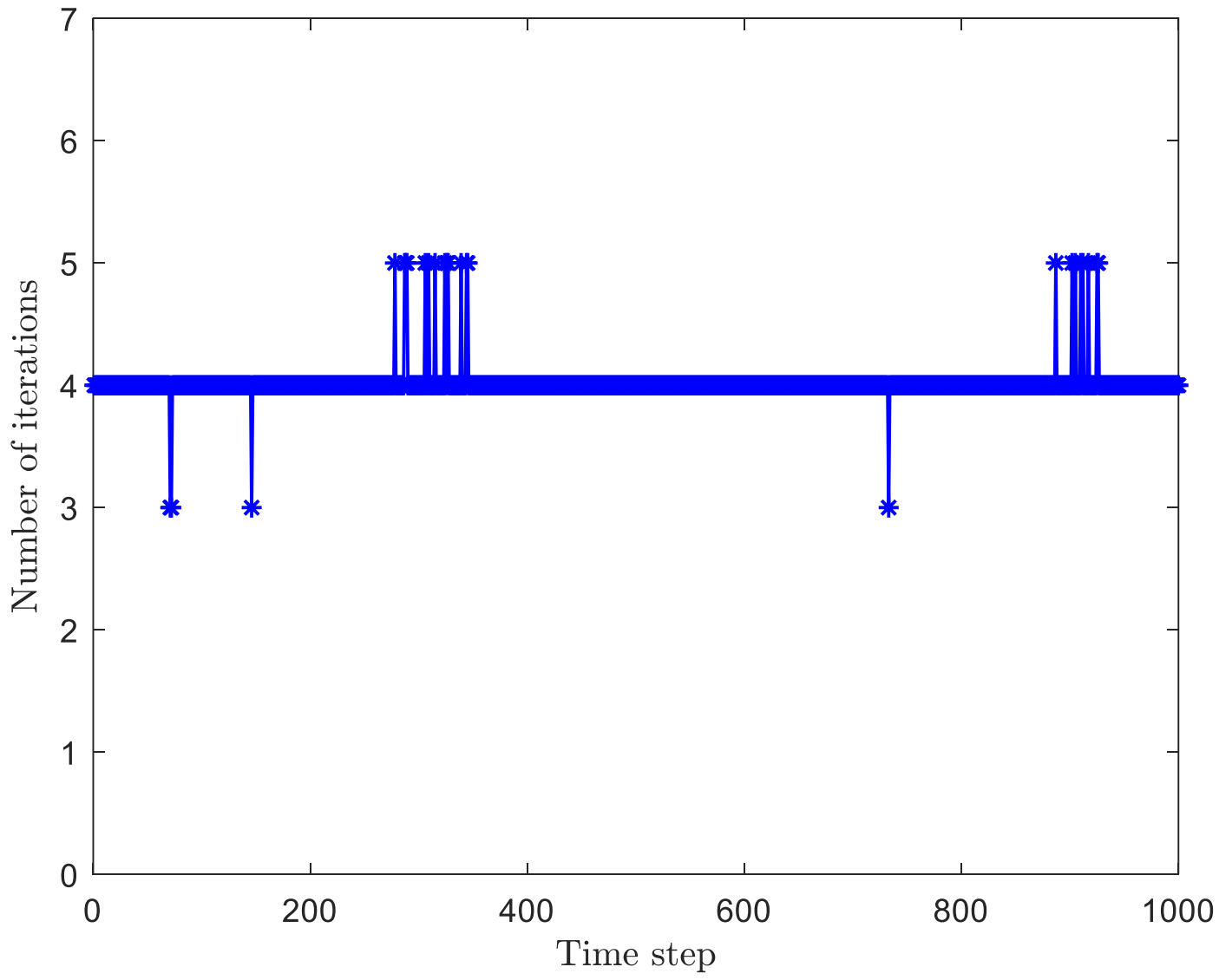

Fig. 8 Number of iterations at each time step.

To test the efficiency of the proposed approach, we also conduct analyses on finite element models with dense mesh grid density. As shown in the right of Fig. 5, three models are considered: (A) 80 elements, 567 DOFs, (B) 1250 elements, 5508 DOFs, (C) 10000 elements, 36663 DOFs. Only the numerical performance of the proposed CR and the full CR approaches are compared as they are both implemented in in-house code and can be profiled in detail. All the other parameters are the same with those of the proceeding analyses. **Table 1** summarizes the total CPU time, as well as the CPU time of the key processes for the two methods. For the Full CR analysis, it is expected that calculating the Jacobian matrix and solving the linearized equations are the most two time-consuming processes. The latter process becomes even time-consuming as the system DOFs increases. For the proposed CR framework, the internal force calculation and the back substitution are the most two time-consuming processes. The CPU times for decomposition of the constant dynamical stiffness matrix $\overline{\boldsymbol{C}}$ is inappreciable, since the decomposition only needs to be taken once during the whole integration process. The back substitution processes, which is believed to be very efficient in ordinary nonlinear implicit analysis, now becomes the bottleneck of the proposed CR framework as the system DOFs increases. The efficiency of this process can be further improved in combination with some advanced techniques, such as GPU-accelerated approach. Our preliminary attempts show that simply using gpuArray for the back substitution processes in the current Matlab environment can achieve a computing speed that is 10x+ faster (for Model C, with NVIDIA RTX3070 GPU device) than the time listed in Table 1. Other technique to accelerate the back substitution processes includes using parallel computing or some advanced sparse triangular matrix solving algorithms (Anzt et al., 2016; Dufrechou and Ezzatti, 2018; Lu et al., 2020). These are topics of our future works. In summary, the existing outcomes show that the proposed CR framework can achieve a very fast computational efficiency compared with that of the full CR method. Besides, the proposed CR framework is promising to be further accelerated easily.

**Table 1** CPU times for the Full CR and the proposed CR methods (unit: second).

| Method | Process | Model A | Model B | Model C |
|---|---|---|---|---|
| Full CR | $\boldsymbol{f}_{\mathrm{INT}}$ calculation | 3.87 | 54.36 | 369.28 |
| | $\boldsymbol{J}$ calculation | 8.57 | 154.59 | 963.56 |
| | $\boldsymbol{J}\Delta\boldsymbol{x}=\boldsymbol{G}_{\mathrm{RES}}$ solving | 10.21 | 233.32 | 3676.92 |
| | Total CPU | 27.46 | 467.52 | 5206.83 |
| Proposed CR | $\boldsymbol{f}_{\mathrm{INT}}$ calculation | 0.37 | 2.03 | 15.29 |
| | Back substitution | 0.21 | 7.32 | 123.78 |
| | $\overline{\boldsymbol{C}}$ decomposition | 0.00 | 0.02 | 0.43 |
| | Total CPU | 0.71 | 11.45 | 149.97 |

6.2 Spinning top

The second example involves a spinning top. This is a famous problem documented in the monograph of classical mechanics, and is well studied if the top is assumed as a rigid body (Krenk and Nielsen, 2014). Here we consider the flexible effects. As shown in Fig. 9, the top is subjected to a gravitational field with the tip being fixed at the origin of the global coordinate system. It rotates with an initial angular velocity $\omega$=100 rad/s along the axis of symmetry. We use the finite element method to carry out the analysis and the model is meshed with 432 eight-node solid elements resulting in 1677 DOFs, as plotted in the right of Fig. 9. The analyses are conducted with a fixed time step size: $h$ = 0.001s, and total simulation time is set as 2 s. The convergence tolerance of the Newton-Raphson iteration is set as $5\times10^{-3}$. For the proposed CR method, the local coordinate system of the top is construed using points A-O-B.

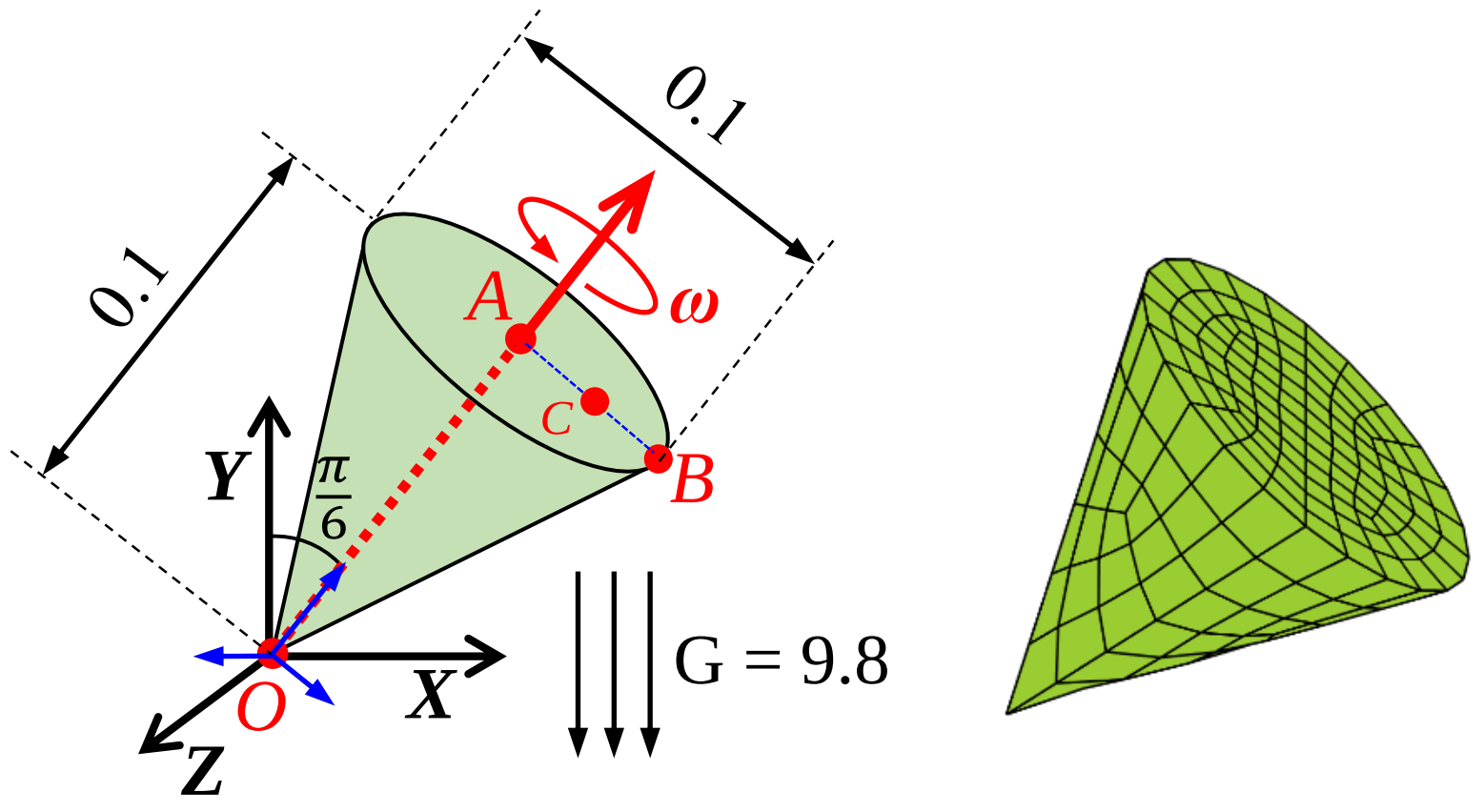

Fig. 9 Spinning top (left), and the finite element model (right).

Fig. 10 presents the time history curves of the global displacements of points A and B. It is found that the results obtained by proposed CR method agree well with those by the other two methods. Due to the high-speed spinning effects, the top moves periodically along the gravity (Y) direction. The global displacements of point A on $X$ and $Z$ directions imply the precession of the top. These phenomenon agree well with existing conclusion for rigid top (Krenk and Nielsen, 2014). Fig. 11 gives the time history curves of the local displacements of a particular point C obtained by the three methods. Good agreements are observed. Statistics show that with the proposed CR method the average iterations needed for each time step is around 4, implying a fast convergence speed. The CPU times needed for the full CR and the proposed CR methods are 408.45 and 4.64s, respectively, demonstrating the high effectiveness of the proposed framework.

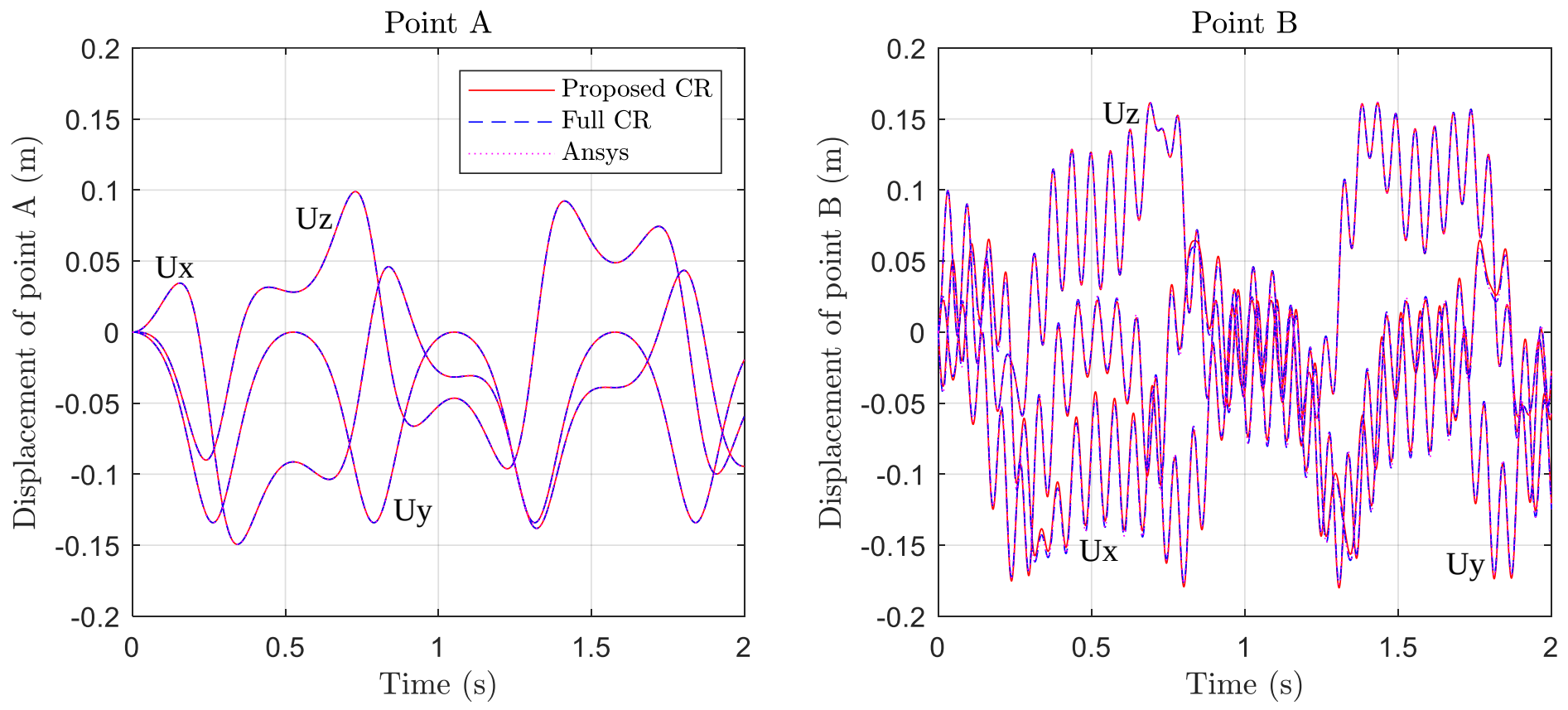

Fig. 10 Time history curves of the global displacements of points A and B in the *X*-*Y*-*Z* directions.

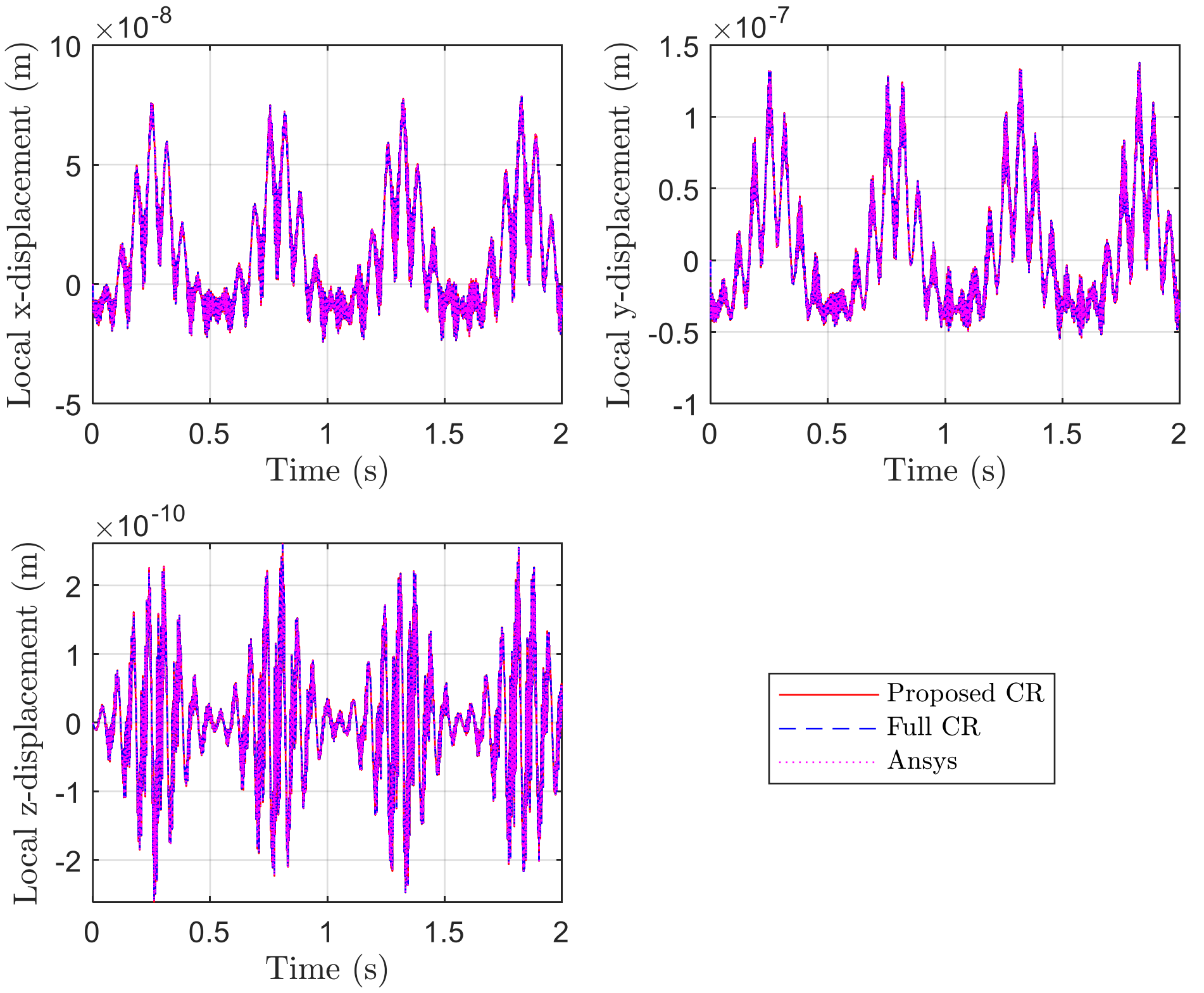

Fig. 11 Time history curves of local displacements of point C in the *x*-*y*-*z* directions.

6.3 A cantilever beam system.

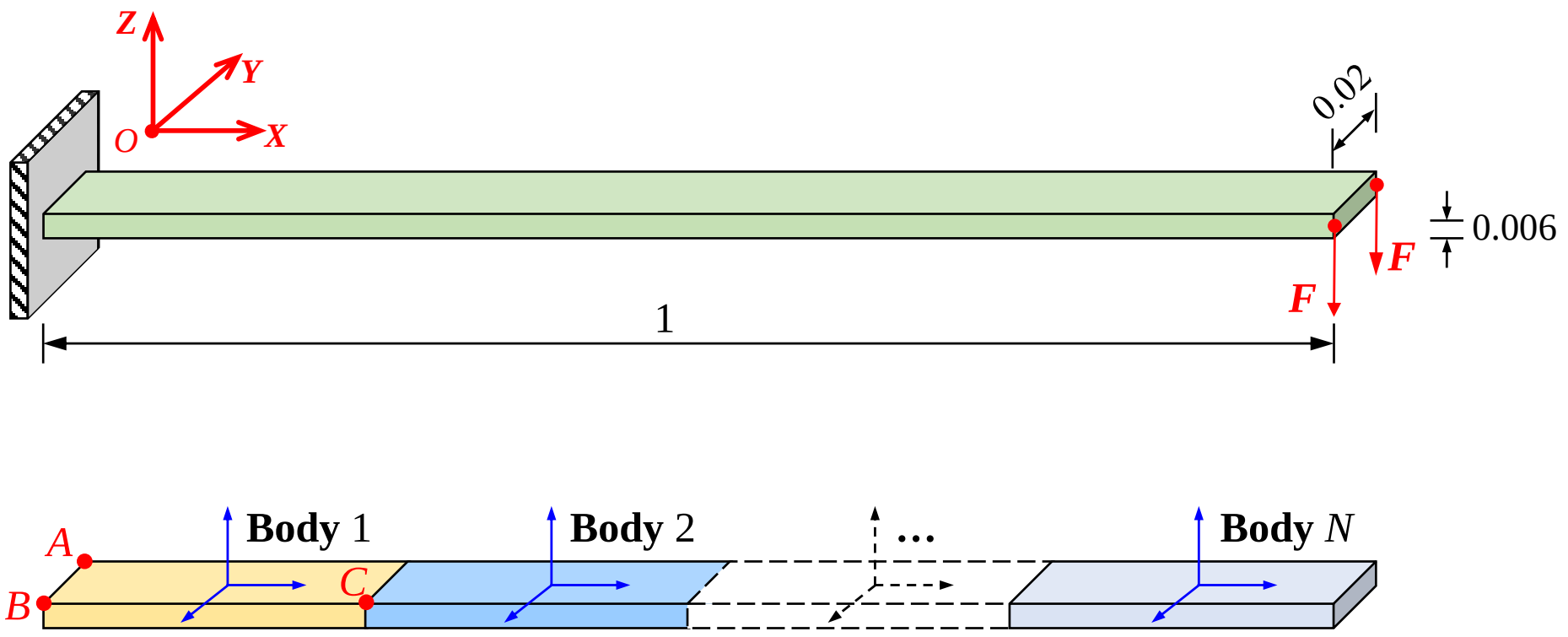

Fig. 12 A cantilever beam system (upper), and the model with the proposed component-level CR framework.

The third example involves static and dynamic analyses of a cantilever beam system. The objective of this examples is to demonstrate the ability of the proposed framework in large deformation analysis. As shown in Fig. 12, the beam is clamped at one end and is subjected to two vertical forces at the corners of the other end. The maximum force vector is set as $F_{max}$ = 30. With this force magnitude, the system produces a downward deflection about the half of the beam length. For the full CR model, the system is meshed with 200×4×1 grid, yielding 800 elements and 6030 DOFs. For the proposed CR framework, due to large deformations, it is clear that one body one component is not feasible. To handle the geometric nonlinearity, we group the finite elements into to $N$ equal sub-structures. Each sub-structure is considered as a flexible body and the rotation is characterized by a single local frame which is calculated based on three corner points (e.g., for Body 1 points A-B-C are used). The adjoined two sub-structures are connected through spherical joints to enforce structural continuity.

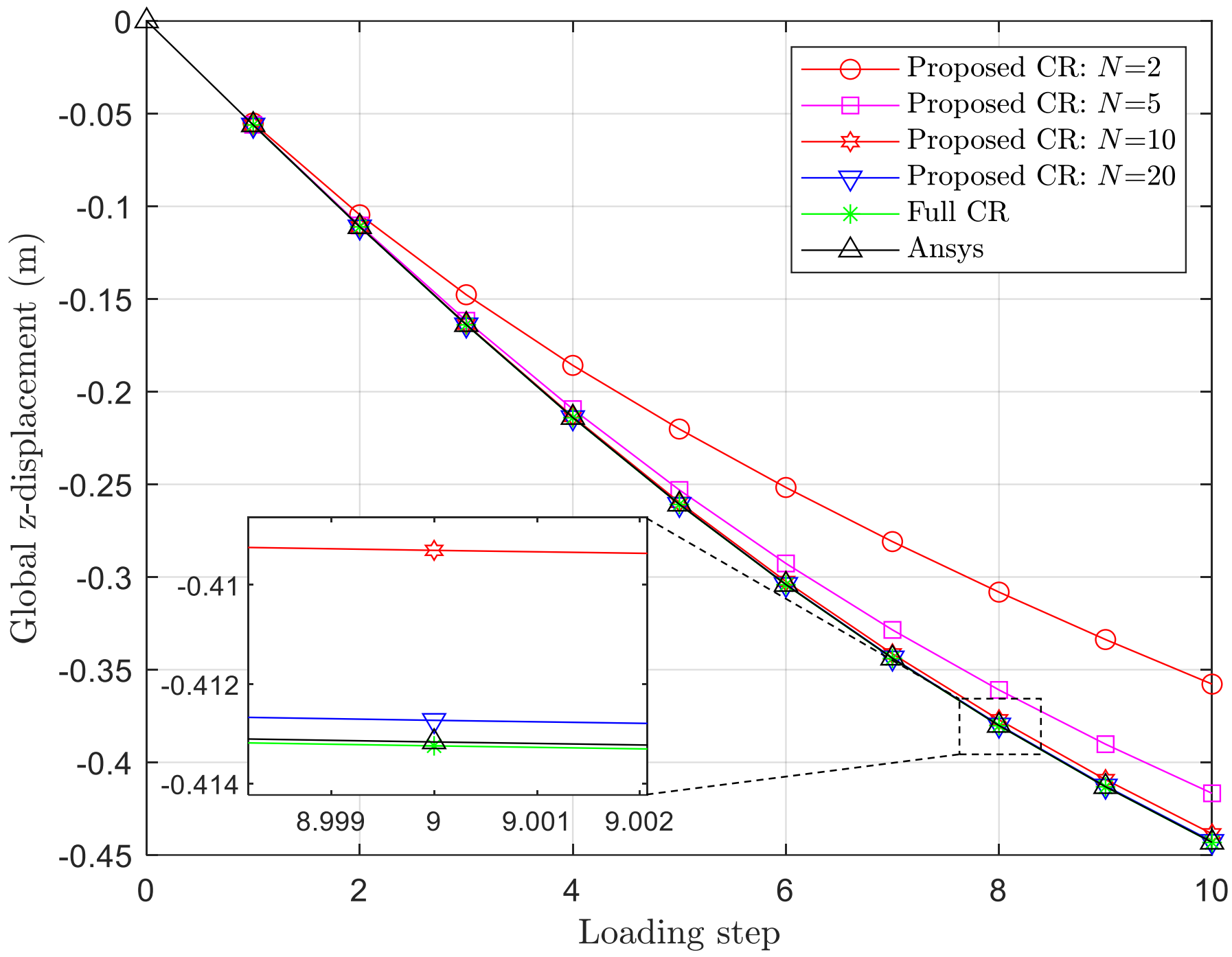

Fig. 13 Vertical displacement of the end point for different models.

We first conduct static analyses. For the proposed CR framework, static results are obtained by the dynamic relaxation approach illustrated in Section 5.2. The damping force is introduced using Eq. (42) with parameter $\zeta$=0.01. Five different cases of sub-structure number are considered: $N$= 2, 5, 10, 20. The results are obtained with 10 equal loading steps. In all analyses the convergence tolerance of Newton-Raphson iteration is set as $5 \times 10^{-4}$. Fig. 13 presents the vertical displacements of the end point for different models. Firstly, it can be observed that the results obtained by the full CR approach agree well with those by Ansys. For the proposed CR approach, the results get convergence to the full CR approach as the sub-structure number increases. This is expected, because the component-level CR model can be considered as an approximation to the full CR model. As previously mentioned, in limit case if each element is treated as a component, the proposed component -based CR model exactly degenerates into the full CR model. For the present analysis, using 20 components is adequate to reflect the geometric nonlinearity. In this situation, the relative error between the results of the proposed CR and those of the full CR method is less than 0.15%.

With same loading conditions we next conduct dynamic analysis. Based on the previous static results, only the case $N$=20 is considered for the proposed CR approach. The total simulation is 0.4s with time step $h$ = 0.002s. The time history curves of the vertical displacement of the end point are shown in Fig. 14. A good agreement is observed. Due to inertial effects, the maximum displacement is about twice as the static results. Some snapshots of the deformed configurations are plotted in Fig. 15 to highlight the large deformations. The CPU times needed for the full CR and the proposed CR methods are 54.64 and 3.53s, respectively, showing the effectiveness of the proposed framework.

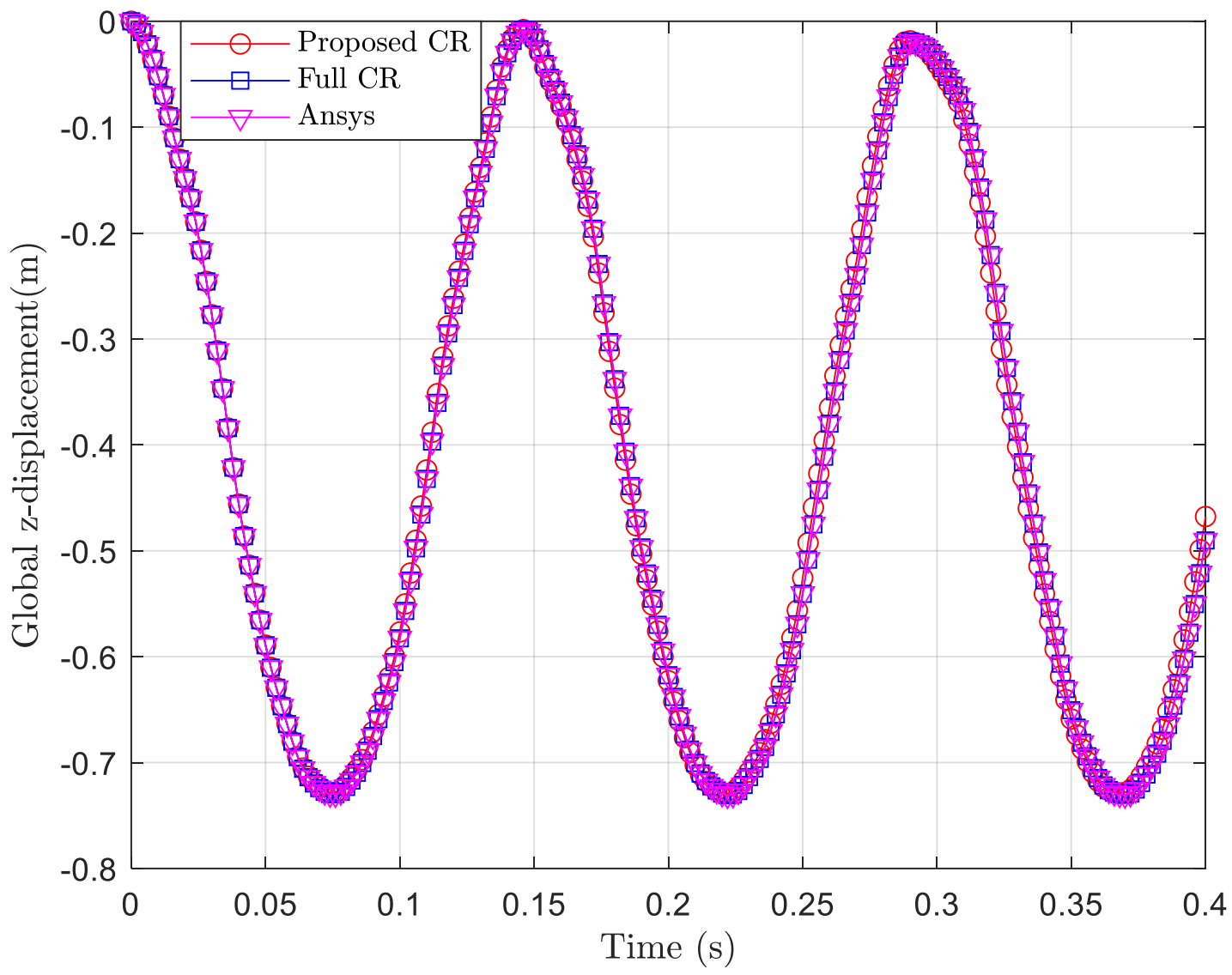

Fig. 14 Time history curves of the vertical displacement of the end point.

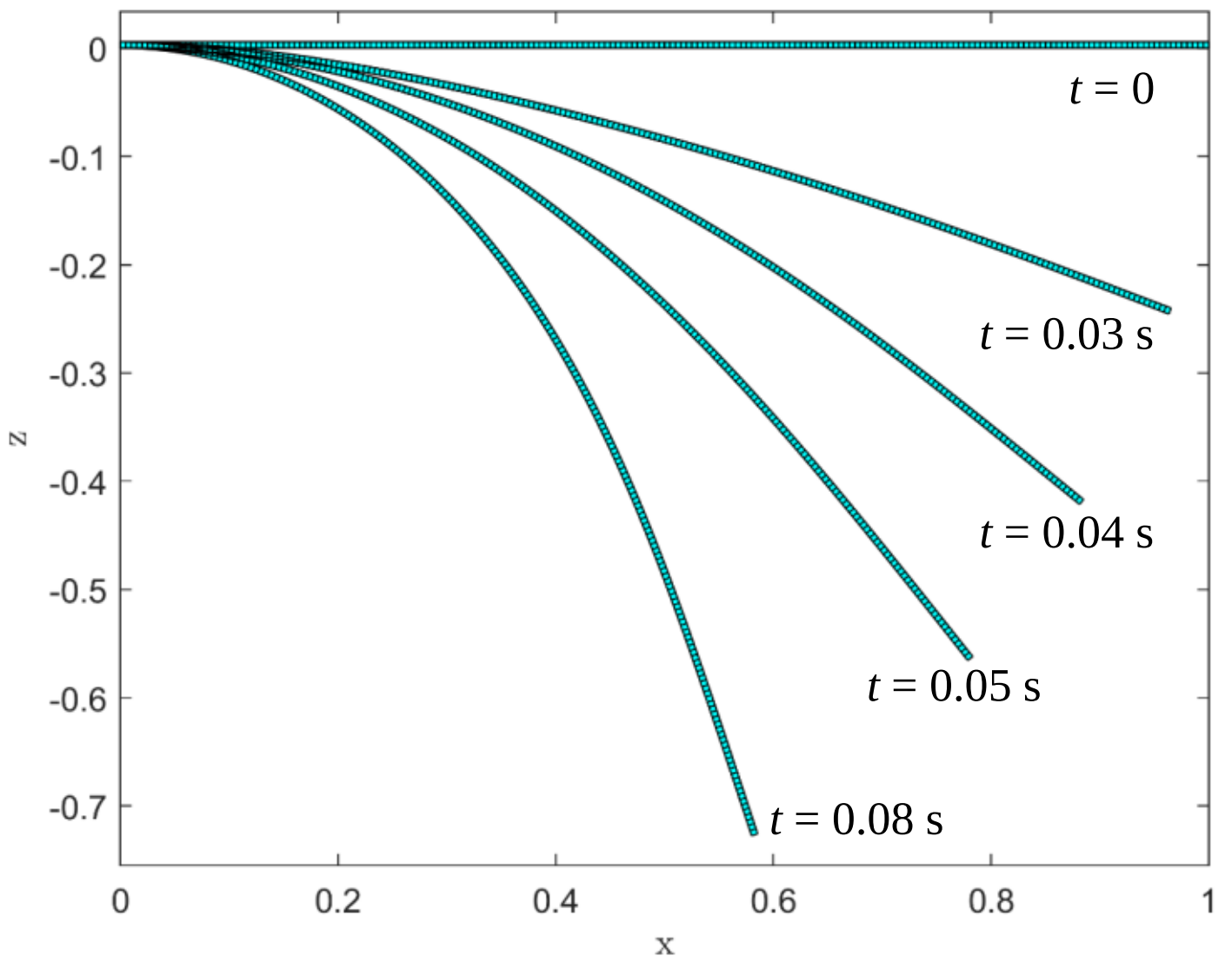

Fig. 15 Some snapshots of the deformed configurations.

## 6.4 Slider-crank mechanism

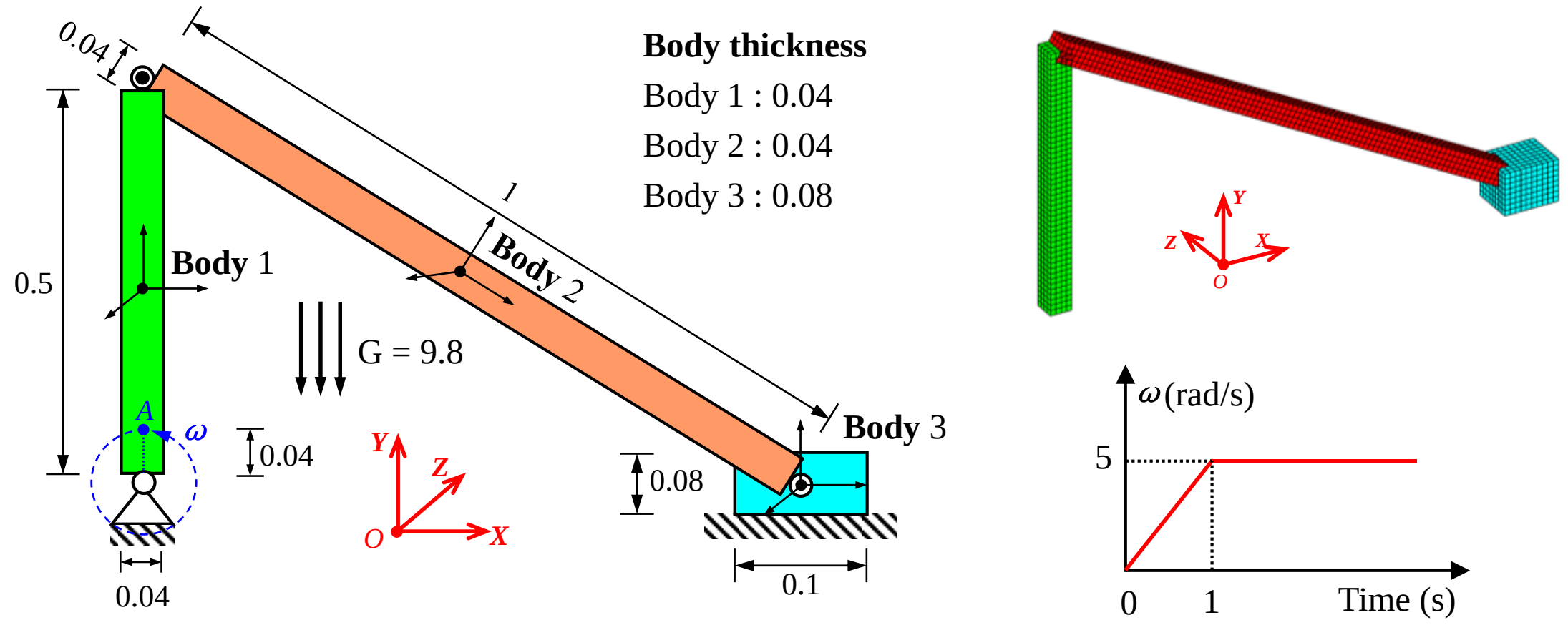

Fig. 16 A slider-crank mechanism (left), and it's finite element model (upper right); the applied time-varying angular velocity (bottom right).

The last example involves a slider-crank mechanism. As shown in Fig. 16, the system consists of three bodies, two links and a mass block. In conventional analysis of this kind of system, to save computational efforts the links are typically modeled with beam elements. With the proposed framework in hand, a different approach is available while maintaining efficiency. For modeling convenience, all the bodies are mesh with solid elements with element size 0.01, yielding a finite element model with 3040 elements and 14073 DOFs shown in the upper right of Fig. 16. The original revolute joints between two adjusted bodies are modeled through spherical joints of the overlapping nodes. The connections between bodies 1 and 3 with the ground can be naturally handled as boundary conditions as in conventional finite element analysis. A motion is applied on the bottom of body 1 to drive system movement, and it is modeled using displacement constraints of the point located at the bottom of the body 1 (point A). The time-varying angular velocity is plotted in the bottom right of Fig. 16. The dynamic analyses are conducted with a fixed time step size: $h$ = 0.01s, and total simulation time 10s. The convergence tolerance of the Newton-Raphson iteration is set as $5\times10^{-5}$. For the proposed CR method, the local coordinate systems of the bodies are determined based on the positions of respective three corners.

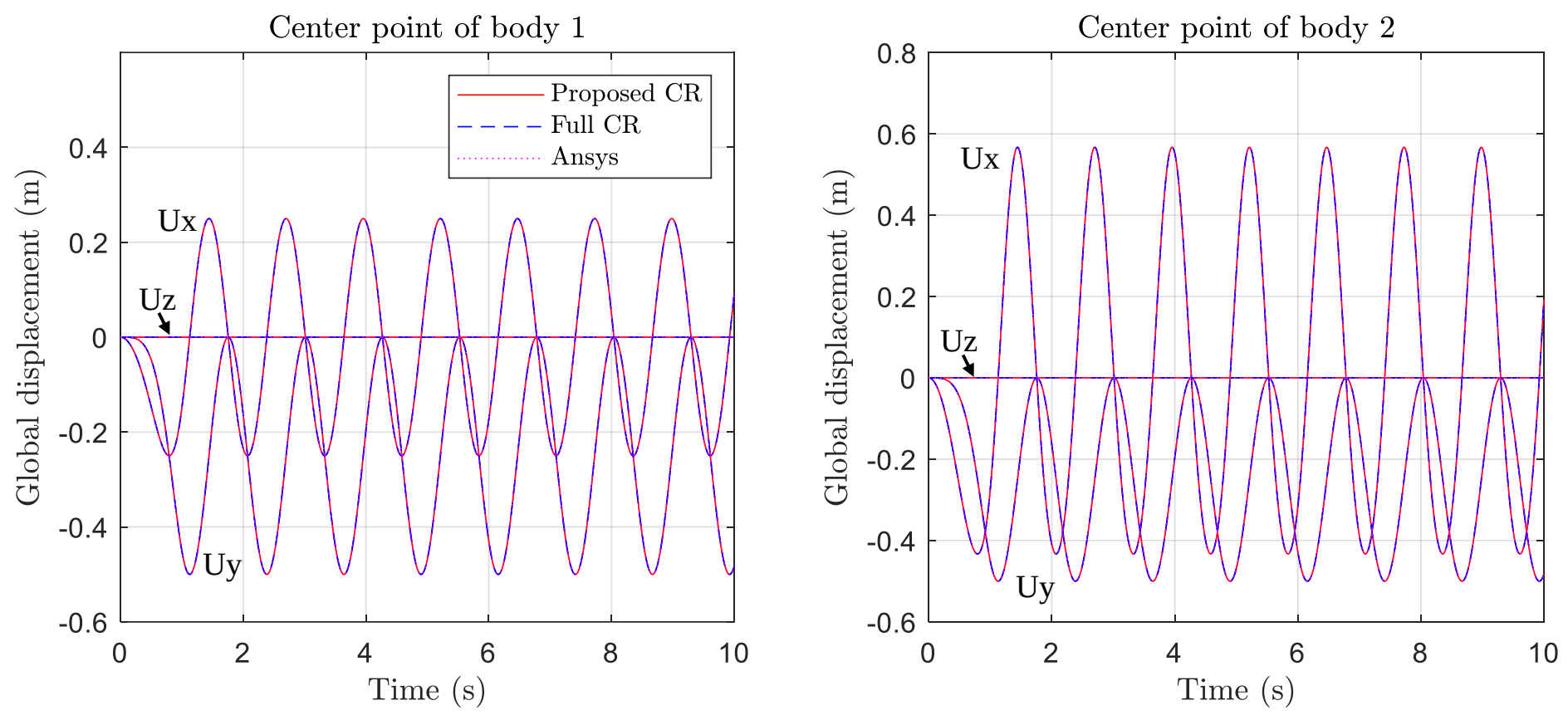

Fig. 17 Time history curves of the center points of the two links.

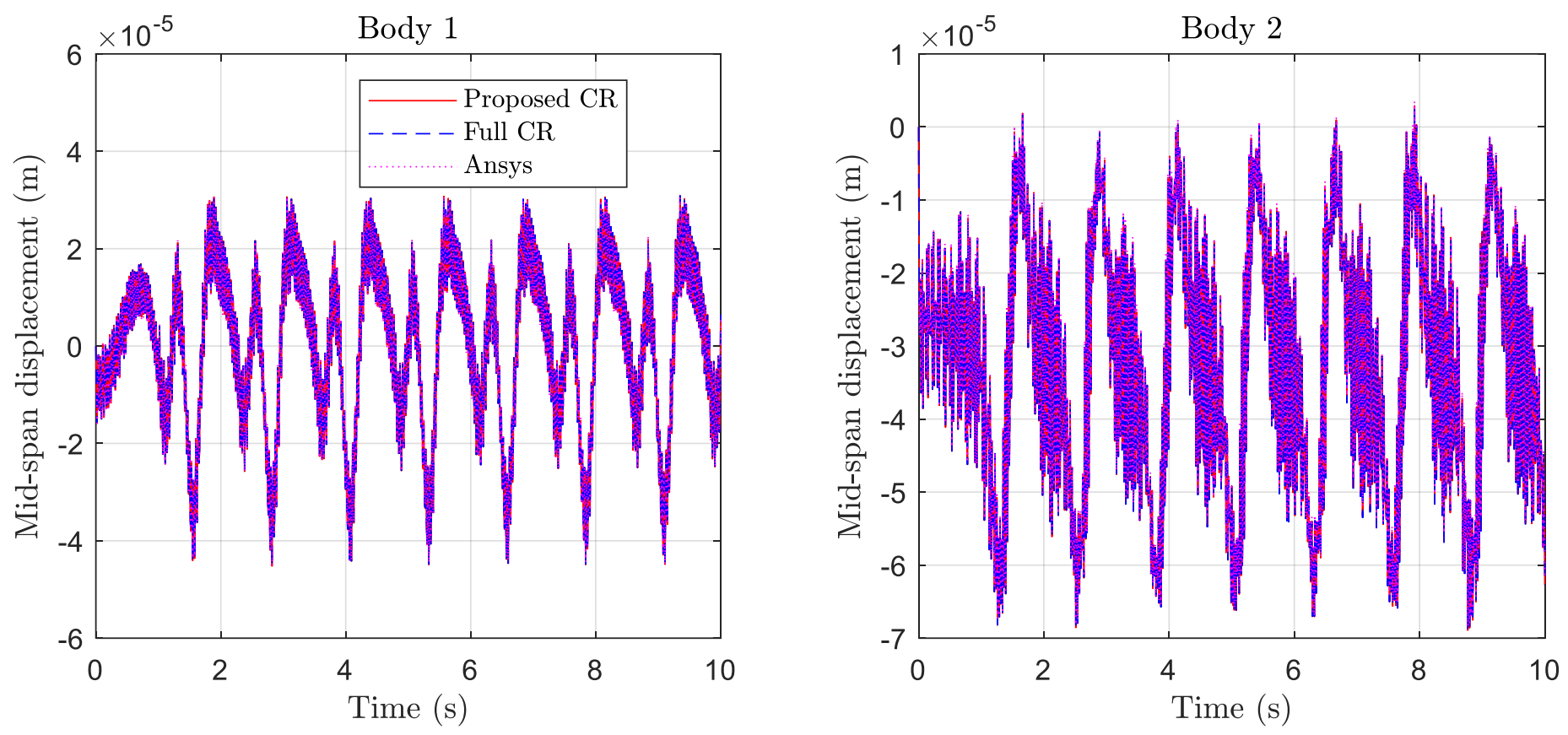

Fig. 18 Time history curves of the mid-span deformation of the two links.

Fig. 17 presents the time history curves of the center points of the two links. The three different approaches yield identical results. These curves demonstrate the large overall motions of the system. Due to the applied motion, the system evolves periodically. The time history curves of the mid-span deformation of the two links are plotted in Fig. 18. Good agreements are observed. To gain an insight into the system movement, Fig. 19 gives some representative configurations of the system. In order to more clearly show the deformation modes of the links, the body deformations are plotted with a magnification of 3000. The color bar reflects the magnitude of the local deformation. The CPU times using for the full CR and the proposed CR methods are 1085.47 and 29.34 s, respectively, demonstrating the high effectiveness of the proposed framework.

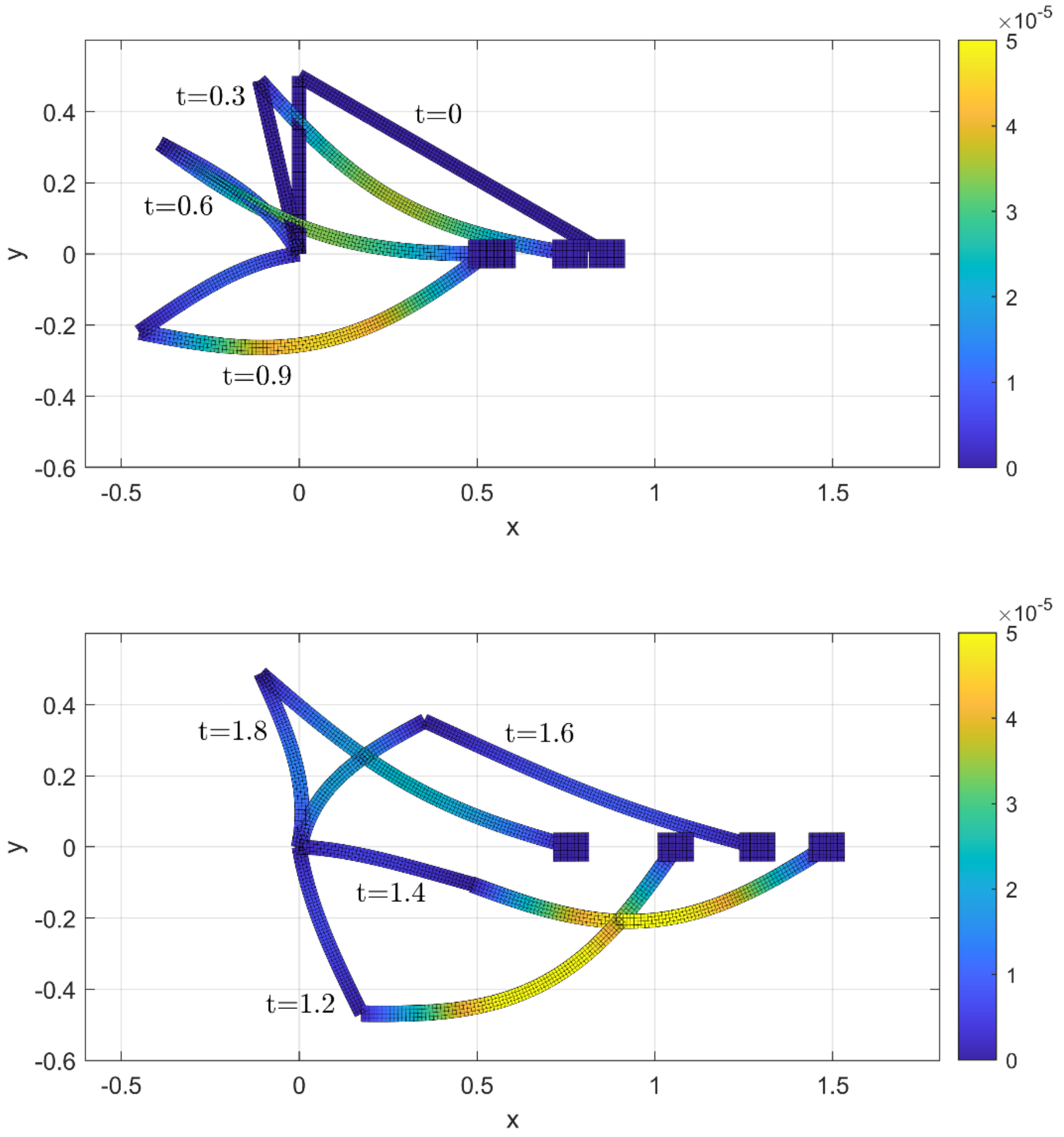

Fig. 19 Some representative configurations of the system.

## 7 Conclusion

This paper proposed a systematic and novel framework, namely component-level CR approach, for upgrading existing 3D continuum finite elements to flexible multibody analysis. A significant innovation of the proposed framework is its high efficient for implicit analysis even with a relatively large number of 3D continuum finite elements and without using any model reduction techniques. The high efficiency is achieved through sophisticated operations in both modeling and numerical implementation phrases. In modeling phrase, as in conventional 3D nonlinear finite element analysis, the nodal absolute coordinates are used as the system generalized coordinates, therefore simple formulations of inertia force terms can be obtained. For elastic force terms, inspired by existing FFRF and conventional CR formulation, a component-level CR modeling strategy is developed. In the numerical implementation phrase, by in combination with Schur complement theory and fully exploring the nature of the proposed component-level CR modeling method, an extremely efficient procedure is developed, which enables us to transform the linearized equations raised from each Newton-Raphson iteration into linear systems where the coefficient matrix remains constant. The coefficient matrix thus can be pre-calculated and decomposed only once, and at all the subsequent time steps only back substitutions are needed, which avoids frequently updating the Jacobian matrix and avoids directly solving the large-scale linearized equation in each iteration. Besides, the back substitutions can be implemented at the component-level with a paralleled manner.

Multiple examples are presented to demonstrate the performance of the proposed framework. It is found that the results show good agreements with those obtained by the conventional CR analysis as well as by the commercial software Ansys. For the numerical examples tested in this work, the outcomes imply that with the proposed framework, the computational efficiency can achieve one or two orders of magnitude faster than the conventional full CR analysis. It is also found that the back substitution process, which is believed to be very efficient in ordinary nonlinear implicit dynamic analysis, now becomes the bottleneck for the proposed framework as the system DOFs increases. The efficiency of this process can be further improved using some advanced techniques, such as GPU-accelerated techniques (Serban et al., 2015) in combination with some advanced sparse triangular matrix solving algorithms, such as presented in (Anzt et al., 2016; Dufrechou and Ezzatti, 2018; Lu et al., 2020). Due to the high efficiency, the proposed framework is also promising to be used as a tool for real time control of flexible multibody system. All these are topics of our future works.

**Acknowledgements**

The authors are grateful for the financial support of the National Natural Science Foundation of China (12372040, 12002072)

## Appendix

### 1. Definition of spin operation

The operation "spin" in Eq. (8) is related to the cross-product and takes the following form for a 3×1 vector $\boldsymbol{r} = [r_1, r_2, r_3]^{\mathrm{T}}$

$$\operatorname{spin}(\boldsymbol{r}) = \boldsymbol{r}\times = \begin{bmatrix} 0 & -r_3 & r_2 \\ r_3 & 0 & -r_1 \\ -r_2 & r_1 & 0 \end{bmatrix} = -\operatorname{spin}(\boldsymbol{r})^{\mathrm{T}} \tag{50}$$

### 2. Derivation of Eq. (7)

Eq. (7) can be obtained following the methodology of Rankin and Nour-Omid (Nour-Omid and Rankin, 1991; Rankin and Nour-Omid, 1988).

First, the term involving the variation of the orthogonal matrix can be written at the node level, as

$$\left.\frac{\partial\left(\operatorname{diag}(\boldsymbol{R})^{\mathrm{T}}\boldsymbol{V}\right)}{\partial\boldsymbol{x}}\right|_{\boldsymbol{V}=\boldsymbol{x}}=\frac{\partial\left(\operatorname{diag}\left(\boldsymbol{R}^{\mathrm{T}}\right)\boldsymbol{x}\right)}{\partial\boldsymbol{u}}=\begin{bmatrix}\frac{\partial\left(\boldsymbol{R}^{\mathrm{T}}\boldsymbol{x}_1\right)}{\partial\boldsymbol{u}}\\ \vdots\\ \frac{\partial\left(\boldsymbol{R}^{\mathrm{T}}\boldsymbol{x}_N\right)}{\partial\boldsymbol{u}}\end{bmatrix} \tag{51}$$

where $\boldsymbol{x}_i\,(1\le i\le N)$ means the 3×1 position vector of the $i$th node. Each component in (51) can be derived using the differential relationship

$$\mathrm{d}\boldsymbol{R}^{\mathrm{T}}\boldsymbol{x}_i=\frac{\partial\left(\boldsymbol{R}^{\mathrm{T}}\boldsymbol{x}_1\right)}{\partial\boldsymbol{u}}\mathrm{d}\boldsymbol{u} \tag{52}$$

By noting that the orthogonal matrix $\boldsymbol{R}$ can be expressed using a skew-symmetric $\boldsymbol{\Omega}$ or its axial vector $\boldsymbol{\omega}$, i.e., $\boldsymbol{R}=\exp(\boldsymbol{\Omega})=\exp(\operatorname{spin}(\boldsymbol{\omega}))$, the variation of the orthogonal matrix thus yields

$$\mathrm{d}\boldsymbol{R}=\operatorname{spin}(\mathrm{d}\boldsymbol{\omega})\boldsymbol{R} \tag{53}$$

where $\operatorname{spin}(\mathrm{d}\boldsymbol{\omega})$ means the infinitesimal rotations of the base vectors of the orthogonal matrix. Noting the relationships $\operatorname{spin}(\mathrm{d}\boldsymbol{\omega})^{\mathrm{T}}=-\operatorname{spin}(\mathrm{d}\boldsymbol{\omega})$ and $\operatorname{spin}(\mathrm{d}\bar{\boldsymbol{\omega}})=\boldsymbol{R}^{\mathrm{T}}\operatorname{spin}(\mathrm{d}\boldsymbol{\omega})\boldsymbol{R}$ hold, for a particular node, we have

$$\begin{aligned}\mathrm{d}\boldsymbol{R}^{\mathrm{T}}\boldsymbol{x}_i&=-\boldsymbol{R}^{\mathrm{T}}\operatorname{spin}(\mathrm{d}\boldsymbol{\omega})\boldsymbol{x}_i=-\operatorname{spin}(\mathrm{d}\bar{\boldsymbol{\omega}})\bar{\boldsymbol{x}}_i\\&=\operatorname{spin}(\bar{\boldsymbol{x}}_i)\mathrm{d}\bar{\boldsymbol{\omega}}=\operatorname{spin}(\bar{\boldsymbol{x}}_i)\sum_{j=1}^{N}\frac{\partial\bar{\boldsymbol{\omega}}}{\partial\bar{\boldsymbol{u}}_j}\frac{\partial\bar{\boldsymbol{u}}_j}{\partial\boldsymbol{u}_j}\mathrm{d}\boldsymbol{u}_j\\&=-\operatorname{spin}(\bar{\boldsymbol{x}}_i)^{\mathrm{T}}\sum_{j=1}^{N}\frac{\partial\bar{\boldsymbol{\omega}}}{\partial\bar{\boldsymbol{u}}_j}\boldsymbol{R}^{\mathrm{T}}\mathrm{d}\boldsymbol{u}_j\end{aligned} \tag{54}$$

With Eqs. (8) and (9), one can verify that the following relation holds,

$$\left.\frac{\partial\left(\operatorname{diag}(\boldsymbol{R})^{\mathrm{T}}\boldsymbol{V}\right)}{\partial\boldsymbol{x}}\right|_{\boldsymbol{V}=\boldsymbol{x}}=-\bar{\boldsymbol{S}}\bar{\boldsymbol{G}}\operatorname{diag}\left(\boldsymbol{R}^{\mathrm{T}}\right) \tag{55}$$

As a result, Eq. (7) is satisfied.